\documentclass[%
prd,
nofootinbib,
superscriptaddress
]{revtex4}
\usepackage{amsmath}
\usepackage{amssymb}
\usepackage{braket}
\usepackage{bm}
\usepackage{aas_macros}
\usepackage{amsmath}
\usepackage{ulem}
\usepackage[dvipdfmx]{graphicx}
\usepackage{color}
\begin{document}

\title{Weak lensing induced by second-order vector mode}
%======================= Author =======================%
\author{Shohei Saga}
\email{saga.shohei@nagoya-u.jp}
\affiliation{Department of Physics and Astrophysics, Nagoya University, Nagoya-city, Aichi 464-8602, Japan}
\author{Daisuke Yamauchi}
%\email{yamauchi@resceu.s.u-tokyo.ac.jp}
\affiliation{Research Center for the Early Universe, Graduate School of Science, 
The University of Tokyo, Hongo, Bunkyo-ku, Tokyo 113-0033, Japan}
\author{Kiyotomo Ichiki}
%\email{ichiki.kiyotomo@c.mbox.nagoya-u.ac.jp}
\affiliation{Department of Physics and Astrophysics, Nagoya University, Nagoya-city, Aichi 464-8602, Japan}
\affiliation{Kobayashi-Maskawa Institute for the Origin of Particles and the Universe, Nagoya University, Nagoya-city, Aichi 464-8602, Japan}
%======================= Abstract =======================%
\begin{abstract}
The vector mode of cosmological perturbation theory imprints characteristic signals on the weak lensing signals such as curl- and B-modes which are never imprinted by the scalar mode.
However, the vector mode is neglected in the standard first-order cosmological perturbation theory since it only has a decaying mode.
This situation changes if the cosmological perturbation theory is expanded up to second order.
The second-order vector and tensor modes are inevitably induced by the product of the first-order scalar modes.
We study the effect of the second-order vector mode on the weak lensing curl- and B-modes.
We find that the curl-mode induced by the second-order vector mode is comparable to that induced by the primordial gravitational waves with the tensor-to-scalar ratio $r = 0.1$ at $\ell \approx 200$.
In this case, the curl-mode induced by the second-order vector mode dominates at $\ell > 200$.
Furthermore, the B-mode cosmic shear induced by the second-order vector mode dominates on almost all scales.
However, we find that the observational signatures of the second-order vector and tensor modes cannot exceed the expected noise 
of ongoing and upcoming weak lensing measurements.
We conclude that the curl- and B-modes induced by the second-order vector and tensor modes are unlikely to be detected in future experiments.
\end{abstract}
%======================= Make title =======================%
\preprint{RESCEU-9/15}
\maketitle
%-----------------------------------------------------------------------------------section
\section{Introduction}
%Overview
The recent remarkable developments of cosmological observations such as the cosmic microwave background (CMB) or large-scale structure help us to build the standard $\Lambda$CDM cosmology.
The new era of high precision cosmology makes it possible to acquire rich information about the expansion history of the Universe or the features of density perturbations~\cite{2013ApJS..208...19H,Ade:2013kta,Tegmark:2006az,Riess:1998cb}.
It is very important to combine several types of observations to reduce degeneracies between cosmological parameters.
The weak lensing effect is a key observable for revealing the late-time evolution of density perturbations.

%weak lensing
The weak lensing effect can be roughly classified into two observables (for reviews, see e.g., \cite{Lewis:2006fu, Bartelmann:1999yn}).
One is called CMB lensing, which is the gravitational deflection by the foreground large-scale structure.
In CMB experiments, we can measure the deflection angle of CMB photons from observed CMB maps 
through the reconstruction technique~\cite{Okamoto:2003zw,Cooray:2005hm,Hirata:2003ka,2012JCAP...01..007N}.
The CMB lensing signals have been precisely detected by the Planck satellite \cite{Ade:2013tyw} and are available to constrain cosmological parameters.
Next-generation CMB observations are planned \cite{Smith:2008an,2011arXiv1102.2181T}, and CMB lensing will become a more important observable in the near future~\cite{2015PhRvD..91d3531N}.
The other observable is called the cosmic shear, which can be measured by observing deformed galaxy images.
The photons emitted from galaxies are deflected by forward density perturbations, deforming the intrinsic shape of galaxies.
Ongoing and upcoming imaging surveys such as the Dark Energy Survey (DES)~\cite{Abbott:2005bi}, Subaru Hyper Suprime-Cam (HSC)~\cite{HSC:coll}, Square Kilometre Array (SKA)~\cite{Brown:2015ucq}, and Large Synoptic Survey Telescope (LSST)~\cite{2009arXiv0912.0201L}, can provide us with high-precision cosmic shear data.
Thus, the weak lensing survey is becoming a more interesting and active area of measurement.

%SVT, curl and B modes
The first-order cosmological perturbation theory includes three independent modes: scalar, vector, and tensor.
Among them, the scalar mode is the dominant component in our Universe and has been well determined by cosmological observations.
Conversely, the vector and tensor modes are subdominant and have not been observed by current observations.
In particular, the vector mode is often treated as the negligible component 
since it rapidly decays in the standard first-order cosmological perturbation theory with perfect fluids.
Nearly all inflation models predict primordial gravitational waves (PGW).
With a nonvanishing amplitude, namely a nonzero tensor-to-scalar ratio $r$, primordial gravitational waves correspond to the tensor mode.
On the basis of current observations, primordial gravitational waves have the small tensor-to-scalar ratio of $r \lesssim 0.1$~\cite{2013ApJS..208...19H,Ade:2013zuv}.
In the context of scalar, vector, and tensor decompositions, the weak lensing effect also can be associated with each mode.
The deflection angle for CMB lensing can be written in the gradient of the scalar lensing potential (gradient-mode) and the rotation of the pseudoscalar lensing potential (curl-mode).
The deformation of the shape of galaxies is described by the Jacobi map, which can be decomposed into even and odd-parity modes (E- and B-modes, respectively).
The vector and tensor modes, rather than the scalar mode, induce the curl- and B-modes.
Therefore, the weak lensing curl- and B-modes are key observables for exploring subdominant modes.

%Examples (1st order)
Some possible sources for the vector and tensor modes in extensions of the standard $\Lambda$CDM cosmology are available.
The weak lensing induced by the primordial gravitational waves has been well studied~\cite{Dodelson:2003bv,Li:2006si}.
The primordial gravitational waves with $r= O(0.1)$ do not have detectable amplitudes for the curl- and B-modes, even under the assumption of ideal experiments.
Cosmic defects are also possible sources of the vector and tensor modes.
The weak lensing effect induced by cosmic strings has been studied, and weak lensing measurements can constrain parameters related to cosmic defects \cite{2012JCAP...10..030Y,Yamauchi:2013fra}.
However, in the first-order cosmological perturbation theory, the vector and tensor modes must have model parameters, e.g., the tensor-to-scalar ratio or the strength of the cosmic string tension.
The amplitudes of the weak lensing signal induced by the above models depend on the model parameters and the generated weak lensing signal has uncertainties.

%2nd order
In the second-order cosmological perturbation theory, the second-order vector and tensor modes are naturally induced by the product of the first-order scalar modes.
These modes do not have free parameters, since the first-order scalar mode is well determined by current observations.
The secondary CMB polarization anisotropy induced by these modes has been discussed in the literature~\cite{Mollerach:2003nq,Pitrou:2010sn,Beneke:2011kc,Fidler:2014oda,Pettinari:2014iha}.
The application of these modes to the weak lensing is also possible and quite interesting.
The contributions of the second-order vector and tensor modes to the gradient- and E-modes are investigated in \cite{Lu:2008ju,Lu:2007cj,Ananda:2006af,Andrianomena:2014sya}.
As the first-order scalar mode can induce the gradient- and E-modes, the contribution from the second-order vector and tensor modes to the gradient- and E-modes must be smaller than that from the first-order scalar mode.
In Ref.~\cite{Sarkar:2008ii}, the authors estimated the curl- and B-mode signals induced by the second-order tensor mode for the first time.
The effect of the second-order tensor mode on the B-mode signal is comparable with that of the primordial gravitational waves with $r = 0.4$ and dominates on small scales, $10 \lesssim \ell$.
However, the second-order tensor mode tends to have a smaller contribution than the second-order vector mode \cite{Mollerach:2003nq,Lu:2008ju,Lu:2007cj,Ananda:2006af,Andrianomena:2014sya}.
Therefore, the weak lensing signal from the second-order vector mode is expected to exceed that from the second-order tensor mode.

%In this paper,
In this paper, we focus on the weak lensing curl- and B-modes induced by the second-order vector mode.
The weak lensing curl- and B-modes are generated not by the scalar mode but by the vector and tensor modes.
Therefore, the curl- and B-modes are good tracers of the subdominant mode in the current Universe.
As the second-order vector mode must have a larger amplitude than the second-order tensor mode, it is important to estimate the weak lensing signal induced by the vector mode.
This paper is organized as follows.
In Sec.~II, we review the second-order cosmological perturbation theory limited to the vector mode.
The standard cosmology comprises the Einstein-Boltzmann system.
We expand the Einstein-Boltzmann system up to the second order and show the evolution and spectrum for the second-order vector metric perturbation.
Furthermore, we discuss the details of the second-order vector metric perturbation and reveal analytic explanations.
In Sec.~III, we summarize the full-sky formalism of CMB lensing and cosmic shear.
In Sec.~IV, we present our main results on the weak lensing signal and offer some discussion concerning detectability.
We assume four survey designs for the cosmic shear measurement.
Finally, we present conclusions in Sec.~V.

Throughout this paper, we use the units in which $c=\hbar = 1$ and a metric signature of $(-,+,+,+)$.
We obey the rule that subscripts and superscripts of greek and latin characters run from 0 to 3 and from 1 to 3, respectively.
%-----------------------------------------------------------------------------------section
\section{Second-order perturbation theory}\label{sec: second-order}
The second-order cosmological perturbation theory is well established in the context of the secondary CMB anisotropy \cite{Hu:1993tc,Senatore:2008vi,Bartolo:2006fj,Bartolo:2005kv,Bartolo:2006cu,Pitrou:2010sn,Pitrou:2008hy,Beneke:2011kc,Beneke:2010eg,Fidler:2014zwa,Saito:2014bxa}.
In this section, we formulate the second-order cosmological perturbation theory, following \cite{Saga:2014jca,Saga:2015bna}.
We work in the Poisson gauge perturbed from the flat Friedmann-Lema\^{\i}tre-Robertson-Walker metric, whose line element is written as
\begin{equation}
{\rm d}s^{2} = a^{2}(\eta)\bigl[ g_{\mu\nu} + \delta g_{\mu\nu}\bigr] {\rm d}x^{\mu}{\rm d}x^{\nu} ~,
\end{equation}
where $\eta$ is the conformal time and $a(\eta )$ denotes the conventional scale factor of a homogeneous
and isotropic universe. 
Moreover we have introduced $g_{\mu\nu}$ and $\delta g_{\mu\nu}$ as the conformal flat four-dimensional metric
and the small metric perturbations, respectively.
Throughout this paper, we adopt the line element in a spherical coordinate system as
\begin{eqnarray}
g_{\mu\nu}{\rm d}x^{\mu}{\rm d}x^{\nu}% &=& -{\rm d}\eta^{2} + \bar{\gamma}_{ij}{\rm d}x^{i}{\rm d}x^{j} \notag \\
&=& -{\rm d}\eta^{2} +{\rm d}\chi^{2} +\chi^{2}\omega_{ab}{\rm d}\theta^{a}{\rm d}\theta^{b} ~, \label{eq: metric spherical}
\end{eqnarray}
where $\chi$ is the comoving distance, and 
$\omega_{ab}{\rm d}\theta^a{\rm d}\theta^b ={\rm d}\theta^2 +\sin^2\theta{\rm d}\varphi^2$ is the metric on the unit sphere.
The perturbed metric in the Poisson gauge can be given by
\begin{eqnarray}
\delta g_{00} &=& -2\Psi~, \\
\delta g_{0i} &=& \sigma_{i}~, \\
\delta g_{ij} &=& -2\Phi\delta_{ij} + h_{ij}~. 
\end{eqnarray}
Under the Poisson gauge, the vector mode $\sigma_{i}$ and the tensor mode $h_{ij}$ obey the transverse condition
$\sigma^i{}_{,i}=h^{ij}{}_{,i}=0$ and traceless condition $h^{i}{}_{i} = 0$ due to the gauge condition.
Here we denote a comma as a spatial derivative and raising or lowering indices is done by $\delta_{ij}$.

The first-order vector mode is usually neglected in the standard $\Lambda$CDM cosmology with perfect fluids since the vector mode is only decaying in the first-order cosmological perturbation theory.
Furthermore, the first-order tensor mode would have a small amplitude, namely, tensor-to-scalar ratio $r\lesssim 0.12$ \cite{Ade:2013zuv}.
Throughout this paper, we therefore neglect the first-order vector and tensor modes.
Note that, in the first-order cosmological perturbation theory, the scalar, vector, and tensor modes evolve independently and it is sufficient to solve the scalar mode only in the first-order theory.
On the other hand, in the second-order cosmological perturbation theory, the scalar, vector, and tensor modes are no longer independent modes.
For instance, the second-order vector and tensor modes are excited from the product of the first-order scalar perturbations even if the first-order vector and tensor modes are absent.
We expand the metric perturbation as $\Psi = \Psi^{(1)}+\frac{1}{2}\Psi^{(2)}$, $\Phi = \Phi^{(1)}+\frac{1}{2}\Phi^{(2)}$, $\sigma_{i} = \frac{1}{2}\sigma^{(2)}_{i}$, and $h_{ij} = \frac{1}{2}h^{(2)}_{ij}$, where we have neglected the first-order vector and tensor modes.
In the following subsections, we will discuss the standard cosmological perturbation theory, which contains the Boltzmann and Einstein equations.

%-----------------------------------------subsection
\subsection{Boltzmann equation}
We consider the Boltzmann equation for the distribution function, $f(x^{\mu}, P^{\mu})$, 
where we denote the canonical momentum ${\rm d}x^\mu /{\rm d}\lambda$ by $P^\mu$ and $\lambda$ is the affine parameter.
In this subsection we present a brief review of the second-order Boltzmann equation, following \cite{Saga:2014jca}.
The distribution function for photons obeys the collisional Boltzmann equation since photons interact with electrons via the Compton scattering.
The collisional Boltzmann equation for photons is written as
\begin{equation}
\frac{{\rm d}f}{{\rm d}\lambda}(x^{\mu},P^{\mu}) = \widetilde{C}[f] ~, \label{eq: Boltzmann eq lambda}
\end{equation}
where $\widetilde{C}[f]$ is the collision term due to the Compton scattering
between photons and electrons in the case of the photon distribution function. 
For massless neutrinos, the collision term in their Boltzmann equation must vanish.
In order to describe the perturbed Boltzmann equation, it is useful to change the coordinate system from the Poisson gauge to the local inertial frame.
Hence we use the the momentum in the local inertial frame $p^\mu =(E,p\hat{\bm n})$ such that $\hat n^i\hat n_i=1$ 
rather than that in the Poisson gauge $P^\mu$, hereafter (see \cite{Senatore:2008vi}).
Here $E$ is the energy measured in the local inertial frame, which obeys the Einstein relation $E^{2} - p^{2} = m^{2}$.
With these variables, we then expand the photon distribution function up to the second order as
\begin{equation}
f(\eta, \bm{x}, p, \hat{\bm n}) = f^{(0)}(\eta, p) + f^{(1)}(\eta, \bm{x}, p, \hat{\bm n}) + \frac{1}{2}f^{(2)}(\eta, \bm{x}, p, \hat{\bm n}) ~.
\end{equation}
The zeroth-order distribution function $f^{(0)}(\eta, p)$ is fixed to the Planck distribution.
It is convenient to define the brightness function $\Delta^{(1,2)}(\eta, \bm{k}, \hat{\bm n})$ in Fourier space as
\begin{equation}
\Delta^{(1,2)} (\eta, \bm{k}, \hat{\bm n})=\frac{\int{dp~p^{3}f^{(1,2)}}(\eta, \bm{k}, p, \hat{\bm n})}{\int{dp~p^{3}f^{(0)}}(\eta, p)}~,
\end{equation}
where we have translated from real space to Fourier space.
Furthermore, we expand the brightness function using spherical harmonics to eliminate the angular dependence:
\begin{equation}
\Delta^{(1,2)}(\eta, \bm{k},\hat{\bm n})=\sum_{\ell}\sum^{\ell}_{m=-\ell}\Delta^{(1,2)}_{\ell,m}(\eta, \bm{k})(-i)^{\ell}\sqrt{\frac{4\pi}{2\ell +1}}Y_{\ell,m}(\hat{\bm n})~.
\end{equation}
The brightness function is related to the temperature perturbation of the photons \cite{2009JCAP...05..014N}.
Here, the multipole coefficient $\Delta_{\ell, m}$ obeys the hierarchical Boltzmann equation
\begin{equation}
\dot{\Delta}^{(1,2)}_{\ell,m}+k\left[ \frac{c_{\ell +1, m}}{2\ell +3}\Delta^{(1,2)}_{\ell +1, m}-\frac{c_{\ell,m}}{2\ell -1}\Delta^{(1,2)}_{\ell -1, m}\right]
=\mathcal{C}^{(1,2)}_{\ell, m}(\eta ,\bm{k} ) +\mathcal{G}^{(1,2)}_{\ell, m}(\eta ,\bm{k})~, \label{eq: Boltzmann eq 1}
\end{equation}
where a dot denotes a derivative with respect to the conformal time, $c_{\ell, m} = \sqrt{\ell^{2}-m^{2}}$,
and $\mathcal{C}^{(1,2)}_{\ell, m}$, $\mathcal{G}^{(1,2)}_{\ell, m}$ are the multipole coefficients of the collision and gravitational terms, respectively.
The multipole coefficient of the collision term corresponds to the right-hand-side in Eq.~(\ref{eq: Boltzmann eq lambda}),
whereas the gravitational term comes from the left-hand-side in Eq.~(\ref{eq: Boltzmann eq lambda}), i.e., the perturbed geodesic equation for photons.
These explicit forms are written in \cite{Saga:2014jca}.
Although in the first-order perturbations the scalar ($m=0$), vector ($m=\pm 1$), and tensor ($m=\pm 2$) modes are completely 
decoupled each other, the vector and/or tensor modes can be generated due to the nonlinear interactions of the scalar ones 
when the second-order perturbations are taken into account.
Note that the multipole coefficient of the neutrino distribution function $\mathcal{N}^{(1,2)}_{\ell, m}$ can be formulated in the same manner without the collision term.

%-----------------------------------------subsection
\subsection{Evolution equation for second-order vector mode}
We denote this subsection to explicitly write down the evolution equation of the second-order vector mode, which 
can be derived through the space-space components of the Einstein equation.
The second-order Einstein tensor, $G^i{}_j$ 
and energy-momentum tensors for the relativistic particles $T_{\rm r}^{i}{}_{j}$ and nonrelativistic particles $T_{\rm m}^{i}{}_{j}$
are given by
\begin{widetext}
\begin{eqnarray}
a^{2}{G^{(2)}}^{i}{}_{j} &=&
2\Phi^{(1)}\left( \Phi^{(1)}{}^{,i}{}_{,j}-\Psi^{(1)}{}^{,i}{}_{,j}\right) + \Phi^{(1)}{}^{,i}\Phi^{(1)}{}_{,j}-\Psi^{(1)}{}^{,i}\Psi^{(1)}{}_{,j}-\left( \Phi^{(1)}{}^{,i}\Psi^{(1)}{}_{,j}+\Phi^{(1)}{}_{,j}\Psi^{(1)}{}^{,i}\right) \notag \\
&&
+\frac{1}{2}\mathcal{H}\left[ 
\dot{h}^{(2)}{}^{i}{}_{j}-\left( \sigma^{(2)}{}^{i}{}_{,j}+\sigma^{(2)}{}_{j}{}^{,i}\right)
\right] 
+\frac{1}{4}\left[ \ddot{h}^{(2)}{}^{i}{}_{j}-h^{(2)}{}^{i}{}_{j}{}^{,k}{}_{,k} -\left( \dot{\sigma}^{(2)}{}^{i}{}_{,j}+\dot{\sigma}^{(2)}{}_{j}{}^{,i}\right)\right] \notag \\
&&+\left(\Phi^{(2)},\Psi^{(2)}\ \mbox{terms}\right) + \mbox{(diagonal part)}~\delta^{i}{}_{j}~, \label{Einstein t}
\end{eqnarray}
\end{widetext}
and
\begin{eqnarray}
T_{\rm r}^{(2)}{}^{i}{}_{j}&=&\rho_{\rm r}^{(0)}\Pi_{\rm r}^{(2)}{}^{i}{}_{j}+\mbox{(diagonal part)}~\delta^{i}{}_{j} ~, \\
T_{\rm m}^{(2)}{}^{i}{}_{j}&=&\rho_{\rm m}^{(0)}v_{{\rm m}0}^{(1)i}v_{{\rm m}0 j}^{(1)}+\mbox{(diagonal part)}~\delta^{i}{}_{j} ~.\label{em t}
\end{eqnarray}
where $\rho,v,$ and $\Pi$ are the energy density, velocity, and anisotropic stress, respectively.
According to the results of the linear perturbations theory, we can split the first-order quantities into
the primordial fluctuations $\phi^{(1)}({\bm k})$ and the transfer functions as
$\Phi^{(1)}(\eta ,{\bm k})=\phi^{(1)}({\bm k})\Phi_{\rm T}(k\eta )$, $\Psi^{(1)}(\eta ,{\bm k})=\phi^{(1)}({\bm k})\Psi_{\rm T}(k\eta )$.
The primordial fluctuations in this paper are assumed to be the random Gaussian field characterized by the primordial power spectrum,
\begin{eqnarray}
\Braket{{\phi^{(1)}}^*({\bm k})\phi^{(1)}({\bm k}')}=(2\pi )^3P_\phi (k)\delta^{3}_{\rm D} ({\bm k}-{\bm k}')~.
\label{eq: primordial power}
\end{eqnarray}
We will further assume a power-law spectrum in the form
\begin{eqnarray}
\frac{k^3}{2\pi^2}P_\phi (k)=\frac{4}{9}\Delta_{\cal R}^2(k_0)\left(\frac{k}{k_0}\right)^{n_{\rm s}-1}~,
\label{eq:primordial power spectrum}
\end{eqnarray}
where $k_0$ denotes the pivot scale.
Hereafter we adopt the scale-invariant spectrum with $\Delta_{\cal R}^2(k_0)=2.4\times 10^{-9}$~\cite{2013ApJS..208...19H} 
and $n_{\rm s}=1$ for simplicity.
The velocity perturbations can be related to the scalar metric potentials through
\begin{eqnarray}
v_{{\rm m}0 i}^{(1)}=-\frac{1}{4\pi Ga^2\rho^{(0)}}\partial_{i}\left(\dot\Phi^{(1)}+{\cal H}\Psi^{(1)}\right) ~.
\label{eq:velocity perturbations}
\end{eqnarray}

In order to describe the evolution equation for the second-order vector mode, it is convenient to expand the variables 
in terms of the mode functions.
To do this, we adopt spherical coordinate basis vectors which are written in the Cartesian coordinate system as
\begin{eqnarray}
\hat{\bm n}&=& (\sin{\theta}\cos{\varphi}, \sin{\theta}\sin{\varphi}, \cos{\theta}) ~, \label{eq: basis vectors1}\\
{\bm e}_{\theta}(\hat{\bm n})&=& (\cos{\theta}\cos{\varphi}, \cos{\theta}\sin{\varphi}, -\sin{\theta}) ~, \label{eq: basis vectors2}\\
{\bm e}_{\varphi}(\hat{\bm n})&=& (-\sin{\theta}\sin{\varphi}, \sin{\theta}\cos{\varphi}, 0) ~, \label{eq: basis vectors3}
\end{eqnarray}
where we have imposed $\epsilon_a^0=0$, ${\bm n}\cdot{\bm\epsilon}_a=0$, and ${\bm e}_a\cdot{\bm e}_b=\omega_{ab}$ in the observer rest frame.
By using these basis vectors, we define the polarization basis as
\begin{equation}
{\bm e}^{(\pm )}(\hat{\bm n}) = {\bm e}_{\theta}(\hat{\bm n}) \pm \frac{i}{\sin{\theta}}{\bm e}_{\varphi}(\hat{\bm n}) ~.
\end{equation}
We then introduce the operator $O^{(\pm 1)}_{i}(\hat{\bm k})$
which can be defined in terms of the polarization vectors in Fourier space, ${\bm e}^{(\pm )}(\hat{\bm k})$,
as (see, e.g., Refs.~\cite{2011PThPh.125..795S,2012arXiv1210.2518S})
\begin{eqnarray}
O^{(\pm 1)}_{i}(\hat{\bm k}) &=& \pm\frac{i}{\sqrt{2}}e^{(\pm )}_i(\hat{\bm k})~, \label{eq: vect basis}
\end{eqnarray}
with $\hat{\bm k}\equiv{\bm k}/k$.
Since these operators satisfy the transverse condition, i.e., $\hat k^iO^{(\pm 1)}_i(\hat{\bm k})=0$,
the vector metric perturbations can be expanded in terms of $O^{(\pm 1)}_i$ as
\begin{eqnarray}
\sigma_{i}(\eta, \bm{x}) &=& \int{\frac{{\rm d}^{3}k}{(2\pi)^3}}\sum_{\lambda = \pm 1}\sigma_{\lambda}(\eta, \bm{k})O^{(\lambda)}_{i}(\hat{\bm k})e^{-i\bm{k}\cdot \bm{x}} ~.
\end{eqnarray}
With these conventions, we can obtain the equation of motion for the second-order vector mode 
by projecting the Einstein equation as
\begin{equation}
\dot{\sigma}^{(2)}_{\lambda}({\bm k})+2\mathcal{H}\sigma^{(2)}_{\lambda}({\bm k}) = {\cal S}^{(2)}_{\lambda}({\bm k}) ~,\label{eq:evolution eq}
\end{equation}
where ${\cal S}^{(2)}_\lambda$ denotes the second-order source terms defined as
\begin{widetext}
\begin{eqnarray}
{\cal S}^{(2)}_\lambda ({\bm k})
&=& \frac{2}{5\sqrt{3}}\frac{1}{k}\left( 8\pi Ga^{2}\rho^{(0)}_{\gamma}\Delta^{(2)}_{2,\lambda}(\bm{k})+8\pi Ga^{2}\rho^{(0)}_{\nu}\mathcal{N}^{(2)}_{2,\lambda}(\bm{k})\right) \notag \\
&& +\int\frac{d^{3}k_{1}}{(2\pi)^{3}}4k_{1}\left[ \Phi^{(1)}(\bm{k}_{1})\Psi^{(1)}(\bm{k}_{2}) \right] \sqrt{\frac{4\pi}{3}}Y^{*}_{1,\lambda}(\hat{\bm k}_{1}) \notag \\
&& -\int\frac{d^{3}k_{1}}{(2\pi)^{3}}\frac{4}{\sqrt{3}}\frac{k^{2}_{1}}{k}\left[ \Phi^{(1)}(\bm{k}_{1})\Phi^{(1)}(\bm{k}_{2})+\Psi^{(1)}(\bm{k}_{1})\Psi^{(1)}(\bm{k}_{2}) \right] \sqrt{\frac{4\pi}{5}}Y^{*}_{2,\lambda}(\hat{\bm k}_{1}) \notag \\
&& +\sum_{{\rm m} = {\rm b}, {\rm dm}}8\pi Ga^{2} \rho^{(0)}_{\rm m}\int{\frac{d^{3}k_{1}}{(2\pi)^{3}}} \left[ \frac{4}{k}v^{(1)}_{{\rm m}0}(\bm{k}_{1})v^{(1)}_{{\rm m}0}(\bm{k}_{2}) \right]\sqrt{\frac{4\pi}{3}}Y^{*}_{1,0}(\hat{\bm k}_{1})\sqrt{\frac{4\pi}{3}}Y^{*}_{1,\lambda}(\hat{\bm k}_{2}) ~, \label{eq:metric}
\end{eqnarray}
\end{widetext}
where $\bm{k} = \bm{k}_{1} + \bm{k}_{2}$,
for simplicity, we omit the time dependence in the above equations, and we have translated from the real space to Fourier space.
We can see that if we consider the linear theory with perfect fluids, namely, without the right-hand-side in Eq.~(\ref{eq:metric}), the vector mode decays with $\propto a^{-2}$.
However, in the second-order cosmological perturbation theory, the nonlinear mode coupling between the first-order scalar modes induces a nondecaying vector mode, which is an entirely second-order effect.
Consequently, the second-order vector mode imprints on the weak lensing.

Before showing the numerical results of the second-order vector mode, we note the features of our numerical calculation \cite{Saga:2014jca,Saga:2015bna}.
To solve the second-order Einstein-Boltzmann system Eqs.~(\ref{eq: Boltzmann eq 1}) and (\ref{eq:evolution eq}), we need the first-order perturbations, e.g., $\Phi^{(1)}$ or $\Delta^{(1)}_{\ell ,m}$ in the Poisson gauge.
We obtain these first-order variables by using the public Boltzmann code such as CAMB \cite{Lewis:1999bs}.
Furthermore, we store the first-order variables in $k$-space from $k_{\rm min} = 5.0\times 10^{-5}~[h{\rm Mpc}^{-1}]$ to $k_{\rm max} = 5.0\times 10^{2}~[h{\rm Mpc}^{-1}]$.
In addition, we solve the Boltzmann hierarchical equation (\ref{eq: Boltzmann eq 1}) at $\ell = 30$ for the first-order Boltzmann equation and at $\ell = 25$ for the second-order Boltzmann equation.
We checked that the results are stable against these choice.

%-----------------------------------------subsection
\subsection{Power spectrum for second-order vector mode}\label{sec: metric}

\begin{figure}[t]
\begin{center}
\rotatebox{0}{\includegraphics[width=0.8\textwidth]{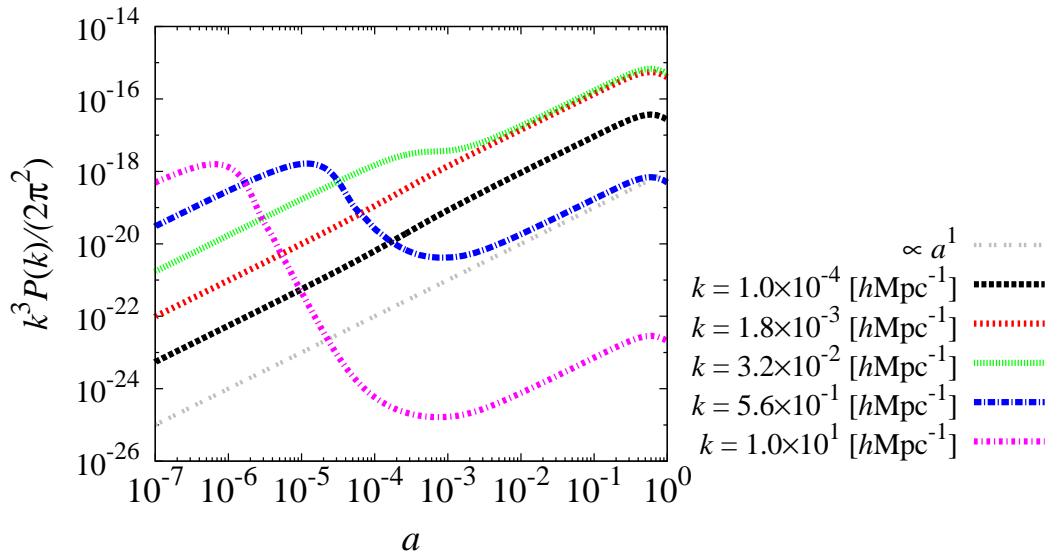}}
\end{center}
\caption{%
The evolution of the second-order vector metric perturbation for scales from $k = 10^{-4}~h{\rm Mpc}^{-1}$ to $10^{1}~h{\rm Mpc}^{-1}$ as indicated in the figures.}
\label{fig: Pk evolution}
\end{figure}

Before evaluating the effect on the weak gravitational lensing, in this subsection
we show the resultant second-order vector mode by performing a fully numerical calculation 
and we then discuss the feature of the power spectrum under some approximations.
To do this, let us define the unequal-time power spectrum for the vector mode as
\begin{equation}
\Braket{\sigma_\lambda^* (\eta ,{\bm k})\sigma_{\lambda'}(\eta' ,{\bm k}')}=(2\pi )^3\delta_{\lambda\lambda'}\delta^{3}_{\rm D}\left({\bm k}-{\bm k}'\right) P_\sigma (\eta ,\eta' ,k)~, 
\end{equation}
where $\Braket{\cdots}$ is the ensemble average.
We derive an expression for the power spectrum by solving the evolution equation.
Equation \eqref{eq:evolution eq} is easily integrated as
\begin{equation}
\sigma^{(2)}_{\lambda}(\eta ,{\bm k}) = \frac{1}{a^{2}(\eta)}\int^{\eta}_{0}d\eta' \left[ a^{2}(\eta'){\cal S}^{(2)}_{\lambda}(\eta' ,{\bm k})\right] ~. \label{eq: sigma integrated}
\end{equation}
Hence we have
\begin{equation}
\Braket{\sigma^{(2)}_{\lambda}(\eta ,{\bm k}){\sigma^{(2)}_{\lambda'}}^* (\eta' ,{\bm k}')}
=\frac{1}{a^2(\eta )a^2(\eta' )}\int^\eta_0{\rm d}\eta_1\int^{\eta'}_0{\rm d}\eta_2 a^2(\eta_1 )a^2(\eta_2 )\Braket{{\cal S}^{(2)}_\lambda (\eta_1 ,{\bm k}){{\cal S}^{(2)}_{\lambda'}}^*(\eta_2 ,{\bm k}')}~.
\label{eq:power spectrum}
\end{equation}
Once we obtain the brightness functions for photons and neutrinos by solving the Boltzmann equation \eqref{eq: Boltzmann eq 1}
and substitute the first-order results for the scalar metric potentials into Eq.~\eqref{eq:metric},
we can obtain the power spectrum for the second-order vector mode though Eq.~\eqref{eq:power spectrum}.
We now solve the evolution equations for the vector mode by performing a fully numerical calculation.
Figure \ref{fig: Pk evolution} shows the equal-time power spectrum for the vector mode induced by the second-order source terms.

\begin{figure}[t]
\begin{center}
\rotatebox{0}{\includegraphics[width=0.7\textwidth]{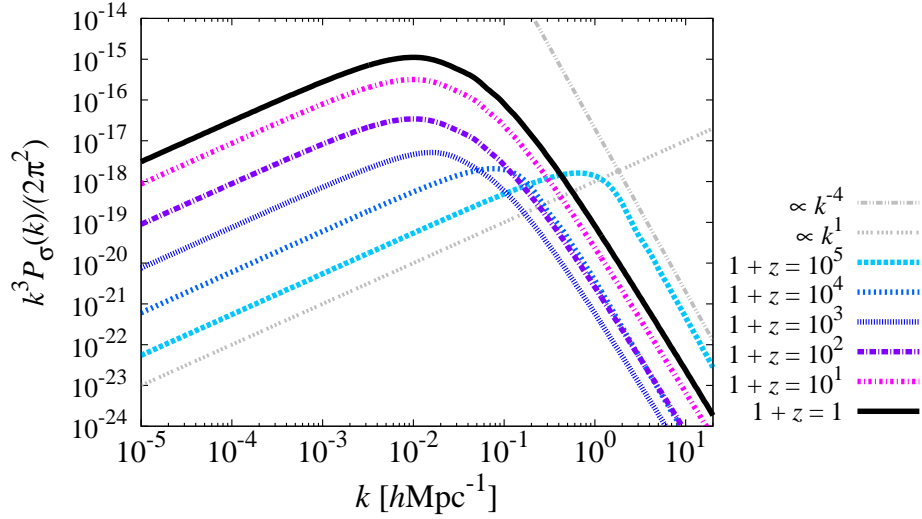}}
\end{center}
\caption{%
The spectra of the second-order vector metric perturbation for redshifts from $1+z = 10^{5}$ to $1$, as indicated in the figure.
Before matter-radiation equality, the feature of the second-order vector metric perturbation was determined by the horizon scale at each time.
On the other hand, after matter-radiation equality, it was determined by the matter-radiation equality scale, namely, $k_{\rm eq} \approx 10^{-2}~h{\rm Mpc}^{-1}$.
We can see that the evolutions are same for all scales after matter-radiation equality $1+z_{\rm eq} \lesssim 3.3\times 10^{3}$.%
}
\label{fig: Pk spectrum}
\end{figure}

For illustrative purposes to show the dependence on the wavenumber, we adopt the wavenumbers as from $k=10^{-4}~h{\rm Mpc}^{-1}$ to $10^{1}~h{\rm Mpc}^{-1}$.
The resultant power spectrum for the second-order vector mode during the radiation-dominated era seems to grow as $\propto a$ on super-horizon scales, 
while it decays on small scales after it enters the horizon scale. 
This is because the source of the second-order vector mode, namely the scalar potential, decays during the radiation-dominated era on sub-horizon scales.
In contrast, during the matter-dominated era it always evolves as $\propto a(\eta )$ for those wavenumbers.
Therefore the second-order vector modes that enter the horizon after the matter-radiation equality time do not undergo the above suppression.
In Fig.~\ref{fig: Pk spectrum}, we plot the dimensionless power spectrum with various values
of the redshift.
We find that it scales as $k^{1}$ on large scales and $k^{-4}$ on small scales and its peak would be determined
by the time of the matter-radiation equality.
%===
%===
%===
%===
We study the analytical description of the power spectrum of the second-order vector mode in the next subsection.

%-----------------------------------------subsection
\subsection{Analytic description of the power spectrum}

In this subsection, we investigate the feature of the power spectrum for the second-order vector mode
analytically.
It is difficult to estimate the second-order vector mode analytically including the purely second-order quadrupole moments for photons and neutrinos.
However, if we assume that the purely second-order quadrupole moments for photons and neutrinos, 
$\Delta^{(2)}_{2,\lambda}$ and $\mathcal{N}^{(2)}_{2,\lambda}$, give negligible contributions, 
the second-order vector metric perturbations are sourced only from the convolution of the first-order scalar metric potentials.
Indeed, numerical computations reveal that the corrections of
the purely second-order quadrupole moments to the vector mode amount to only about $\lesssim O(10^{-3}) \% $ during the matter-dominated era, whereas, the quadrupole moments contribute several tens percent to the vector mode during the radiation-dominated era.
Even in the radiation-dominated era, the scalar potentials are still dominant in Eq.~(\ref{eq:metric}).
In this paper, we focus on the weak lensing signals, which are mainly determined by the contributions
after the matter-radiation equality, in which regime the quadrupole moments contribute at most $O(10^{-3}) \% $.
Therefore, it is sufficient to consider only the scalar metric potentials and we ignore the second-order quadrupole moments, if we give a rough estimation.

To simplify the analysis, we adopt the condition such that the two first-order scalar metric potentials are equal, i.e., $\Phi^{(1)}=\Psi^{(1)}$.
While this is valid only if the first-order quadrupole moments are negligibly small, we keep it just for
a qualitative understanding of the behavior of the power spectrum for the vector mode.
During the radiation-dominated era, the scalar potentials are constant on super-horizon scales while they decay on sub-horizon scales.
On the other hand, they freeze on all scales during the matter-dominated era (e.g., see \cite{Baumann:2007zm}).
We note that under this condition the second term of the right-hand-side in Eq.~\eqref{eq:metric} should vanish.
This is understood as follows:
The condition we impose here implies that ${\bm k}_1$ and ${\bm k}_2$ are interchangeable.
Moreover, the spherical harmonics has the following property:
\begin{equation}
k_{1}Y^{*}_{1,\lambda}(\hat{\bm k}_{1}) + k_{2}Y^{*}_{1,\lambda}(\hat{\bm k}_{2}) = \sqrt{\frac{3}{4\pi}}k\delta_{m,0} ~,
\end{equation}
where we have imposed ${\bm k}={\bm k}_1+{\bm k}_2$.
As a result, the second line in Eq.~\eqref{eq:metric} gives negligible contributions to the vector mode
in the absence of the quadrupole moments.

Let us evaluate the vector mode during the radiation-dominated era.
Since the fourth term in Eq.~\eqref{eq:metric} is estimated through Eq.~\eqref{eq:velocity perturbations}
as $8\pi Ga^2\rho^{(0)}_{\rm m}v_{{\rm m}0}^2\sim (\rho_{\rm m}^{(0)}/\rho^{(0)})\Phi^2$,
it is suppressed by the factor $\rho_{\rm m}^{(0)}/\rho^{(0)}\ll 1$ compared with the third term.
Therefore we found that the third term gives a dominant contribution to the second-order vector mode.
Using the explicit expression for the spherical harmonics,
the power spectrum for the vector mode induced by the third term in Eq.~(\ref{eq:metric}) can be written as
\begin{equation}
\frac{k^{3}}{2\pi^{2}}P_{\sigma} \propto k\int{\rm d}^{3}k_{1}P_\phi (k_{1})P_\phi (k_{2})T^{2}(\eta, k_{1},k_{2})
\left[ k^{4}_{1}\sin^{2}{\theta_{1}}\cos^{2}{\theta_{1}} - k^{2}_{1}k^{2}_{2}\sin{\theta_{1}}\sin{\theta_{2}}\cos{\theta_{1}}\cos{\theta_{2}}\right] ~, \label{eq: SH power1}
\end{equation}
where ${\bm k}_2={\bm k}-{\bm k}_1$, $\hat{\bm k}\cdot\hat{\bm k}_i=\cos\theta_i$, and
the integrated transfer function $T(\eta ,k_{1},k_{2})$ is defined in terms of the transfer functions
for the scalar potential $\Phi_{\rm T}$ as
\begin{equation}
T(\eta , k_{1},k_{2}) = \frac{1}{a^{2}(\eta)}\int_0^\eta{\rm d}\eta'a^{2}(\eta')\Phi_{\rm T}(k_{1}\eta')\Phi_{\rm T}(k_{2}\eta') ~. \label{eq: transfer integral 1}
\end{equation}
With a help of the definition of ${\bm k}_2$ and introducing the direction cosine $\mu_1\equiv\cos\theta_1$, 
Eq.~\eqref{eq: SH power1} can be reduced to 
\begin{equation}
\frac{k^{3}}{2\pi^{2}}P_{\sigma} \propto k\int_0^\infty {\rm d}k_1\int_{-1}^1{\rm d}\mu_1\, k_1^5P_\phi (k_1)P_\phi (k_2)T^{2}(\eta, k_1,k_2)
\left( 2k_1\mu_1 -k\right)\mu_1\left( 1-\mu_1^2\right) ~.
\label{eq:power estimation}
\end{equation}
To perform this integration analytically, we assume that the transfer function of the scalar potential during the radiation-dominated era is approximated as~\cite{Baumann:2007zm}
\begin{equation}
\Phi_{\rm T}(k\eta) = \frac{1}{1 + \left( k\eta\right)^{2}} \quad\left(\eta <\eta_{\rm eq}\right)\,.
\end{equation}
Substituting the above transfer function into Eq.~\eqref{eq: transfer integral 1}, we have
\begin{equation}
T(\eta, k_{1},k_{2}) = \frac{1}{\eta^{2}k_{1}k_{2}\left( k^{2}_{1}-k^{2}_{2}\right) }\Bigl[ k_{1}\arctan\left( k_{2}\eta \right) - k_{2}\arctan\left( k_{1}\eta \right)\Bigr] 
\quad\left(\eta <\eta_{\rm eq}\right) .
\end{equation}
In order to evaluate the behavior of the power spectrum, we split the integral of $k_1$ in Eq.~\eqref{eq:power estimation} into
two parts: the contributions from $k_1>k$ and $k_1<k$ for given $k$.
In the former case, the dummy variables $\bm{k}_{1}$ and $\bm{k}_{2}$ are related through
$k_2=k_1\bigl[ 1-(k/k_1)\mu+O((k/k_1)^2)\bigr]$ and the integrated transfer function can be reduced to the following form:
\begin{equation}
T(\eta, k_{1},k_{2})\approx \frac{1}{2k^{3}_{1}\eta^{2}}\left[ \arctan (k_{1}\eta) - \frac{k_{1}\eta}{1 + (k_{1}\eta)^{2}} \right]
\equiv\eta\,\tau_1 (k_1\eta )\quad\left(\eta <\eta_{\rm eq}\right) ~.
\end{equation}
Hence the contributions from the products of the first-order scalar potentials with their wavelengths shorter than $k$ are
\begin{eqnarray}
\frac{k^{3}}{2\pi^{2}}P_{\sigma}
&\propto& k\int_{k}^\infty{{\rm d}k_{1}}\int_{-1}^{1}{{\rm d}\mu_{1}}\,k_1^6P^2_{\phi}(k_1)\,\eta^2\, \left(\tau_1 (x_1)\right)^2\,\mu_{1}^2(1-\mu^{2}_{1}) \notag \\
&\propto& k\eta \int^{\infty}_{k\eta}{{\rm d}x_{1}}\left(\tau_1 (x_1)\right)^2 
\equiv k\eta\, \beta_1 (k\eta )~,
\label{eq:longer}
\end{eqnarray}
where we have used the scale-invariance of the primordial power spectrum Eq.~\eqref{eq:primordial power spectrum} and 
we have changed the variable $k_1$ to $x_{1} \equiv k_{1}\eta$.
Since $\tau_1 (x)$ behaves as $x^0$ for $x\ll 1$ and $x^{-3}$ for $x\gg 1$, the integral of $\beta_1$ in Eq.~\eqref{eq:longer} 
can be evaluated as a function of $k\eta$ : $\beta_1\propto (k\eta )^0$ for $k\eta\ll 1$, and $\beta_1\propto (k\eta)^{-5}$ for $k\eta\gg 1$.
Substituting this into Eq.~\eqref{eq:longer}, we calculate the contributions from modes with $k_1>k$ in Eq.~\eqref{eq:power estimation}:
\begin{eqnarray}
\frac{k^{3}}{2\pi^{2}}P_{\sigma}(k<k_1,\eta <\eta_{\rm eq})
\propto
		\begin{cases}
			\left( k\eta\right)^{1} & k\eta\ll 1\\
			\left( k\eta\right)^{-4} & k\eta\gg 1 
		\end{cases}
		\,. \label{eq: power analytic}
\end{eqnarray}
We can reproduce the behavior of the power spectrum, namely $\propto k^{1}$ for super-horizon scales and $\propto k^{-4}$
for sub-horizon scales, which can be seen in the numerical calculations.

Following the same manner, we can analyze the opposite case, namely $k_1<k$.
Expanding Eq.~\eqref{eq:power estimation} in terms of the small quantity $k_1/k\ll 1$,
we find that the leading order term vanishes due to the angular integration.
Furthermore, we also find that the-next leading order term is suppressed by the power $k^{3}_{1}$.
Hence the contributions from modes with their wavelengths longer than $k$ 
are suppressed by the factor $k_{1}/k$ and can be treated as subdominant components.
Note that although the above estimations do not work around $k\approx k_{1}$, the results in Eq.~(\ref{eq: power analytic}) is expected to be still correct as long as we estimate the behavior roughly for the following reasons.
The integrand in Eq.~(\ref{eq:power estimation}) does not diverge at $k = k_{1}$.
The estimations on $k < k_{1}$ and $k>k_{1}$ are smoothly connected.
Therefore, the contribution from $k\approx k_{1}$ is at most same order as that from $k < k_{1}$.
Combining these results, we conclude that the power spectrum for the second-order vector mode 
during the radiation-dominated era
is determined by the convolution of the scalar potentials with shorter wavelengths.

We discuss the peak shift of the second-order vector mode shown in Fig.~\ref{fig: Pk spectrum} from $1+z = 10^{5}$ to $10^{4}$.
During the radiation-dominated era, the first-order scalar potential remains constant on super-horizon scales while it decays on sub-horizon scales.
Therefore, the second-order vector mode can grow due to the constant scalar potential on super-horizon scales.
On sub-horizon scales, the second-order vector mode conversely decays due to the decaying scalar potential.
As a result, the peak of the second-order vector mode is determined by the horizon scale at the corresponding era and the peak keeps shifting until the matter-radiation equality.

We next consider the vector mode after the radiation-dominated era.
The evolution during this era can be easily understood through Eq.~(\ref{eq: sigma integrated}).
Generally, when the second-order source term remains constant (i.e., ${\cal S}^{(2)}_{\lambda} = {\rm const.}$), 
the vector mode evolves as $\sigma^{(2)}_{\lambda}\propto \eta^{1}$.
This condition is actually satisfied since the scalar potentials during the matter-dominated era freeze on all scales, as mentioned above.
Hence the evolution of the second-order vector mode is given by
\begin{equation}
\frac{k^{3}}{2\pi^{2}}P_{\sigma} \propto \bigl( \sigma^{(2)}_{\lambda}\bigr)^{2} \propto \eta^{2}\propto a^{1} ~~~\mbox{(for all scales)}~.
\end{equation}
During the matter-dominated era, the shape of the spectrum for the vector mode does not dramatically change since the growing features are the same over all scales.
Therefore, the information about the power spectrum during the radiation-dominated era propagates to one during the matter-dominated era, i.e., 
the dimensionless power spectrum during the matter dominated era is still in proportion to $k^{1}$ for super-horizon scales and $k^{-4}$ for sub-horizon scales,
respectively.
Although the global feature can be understood as above, in more detail
small shifts of the scalar potential such as $\Phi \to 9/10 \Phi$ during the matter-radiation equality 
induce an additional small suppression of the second-order vector mode, as seen in Fig.~\ref{fig: Pk evolution}.
After the universe is dominated by the dark energy, the scalar potentials begin to decay for all scales, implying that
the second-order vector mode generated by these potentials also decays.\\

Before closing this subsection, we introduce the analytic model of the power spectrum, which is originally derived in \cite{Mollerach:2003nq} (hereafter referred to as MHM).
The explicit form of the approximate solution can be written as
\begin{equation}
\frac{k^{3}}{2\pi^{2}}P_\sigma^{(\rm MHM )}(k,\eta ,\eta' ) = \frac{18}{25^{2}}C_{\rm V}\Delta^{4}_{\mathcal{R}}(k_{0})
k^2\left( \frac{k}{k_{*}}\right)^{-1} W_{\rm V}(k/k_{*})F(z)F(z')
 ~, \label{eq: MHM vector}
\end{equation}
where $C_{\rm V} \approx 0.026$, $W_{\rm V}(x) = (1 + 5x+3x^{2})^{-5/2}$, and $k_{*} = \Omega_{{\rm m}0}h^{2}~{\rm Mpc}^{-1}$ with
$\Omega_{{\rm m}0}$ and $h$ being the present cosmological parameter of the nonrelativistic matter and the Hubble constant $H_0$
in unit of $100\ [{\rm km\,s^{-1}\,Mpc^{-1}}]$, respectively.
The function of a redshift $F(z)$ is given by
\begin{equation}
F(z) = \frac{2g^{2}(z)E(z)f(\Omega_{\rm m}(z))}{\Omega_{{\rm m}0}H_{0}(1+z)^{2}} ~,
\end{equation}
where $E(z)=\Omega_{{\rm m}0}(1+z)^{3} +\left( 1-\Omega_{{\rm m}0}\right)$, $\Omega_{\rm m}(z)=\Omega_{{\rm m}0}(1+z)^3/E^2(z)$.
We adopt $\Omega_{{\rm m}0}=0.27$ as the fiducial value.
We denote $f(\Omega_{m}(z))$ and $g(z)$ as the dimensionless linear growth rate and the growth suppression factor, respectively.
One can find that $f$ and $g$ are well approximated as $f\approx\Omega_{\rm m}^{7/4}(z)$ and 
\begin{equation}
g(z) \propto \Omega_{\rm m}(z)\left[ \Omega^{4/7}_{\rm m}(z) -\Omega_{\Lambda}(z) +\left( 1+\Omega_{\rm m}(z)/2 \right) \left( 1+\Omega_{\Lambda}(z)/70\right)\right]^{-1} ~,
\end{equation}
where $\Omega_{\Lambda}(z)=\left( 1-\Omega_{{\rm m}0}\right)/E^2(z)$ and we will normalize $g$ so that $g(0) = 1$ \cite{1991MNRAS.251..128L,1992ARA&A..30..499C,Mollerach:2003nq}.
We find that the transfer functions derived in MHM and those determined by the numerical calculation match after the matter-radiation equality.
However, we should emphasize that 
for the MHM approximate power spectrum the effect from the evolution of the vector mode over all wave numbers during the radiation-dominated era 
is assumed to be neglected.
As we will see in the subsequent analysis, this approximation leads to the non-negligible difference between the full-numerical
and analytic power spectrum.

%-----------------------------------------subsection
\subsection{Tensor mode}

As we mentioned in the Introduction, the curl mode of the CMB lensing and the B-mode shear 
can be generated by not the scalar metric perturbations but the vector and/or tensor metric perturbations.
In this subsection, to compare with the second-order vector mode, tensor modes 
are considered as alternative sources of the observables we focus on.
In particular, we consider primordial gravitational waves and second-order tensor mode as intriguing
examples for tensor metric perturbations.
To describe the spectrum for the tensor mode, we define the spin-$\pm 2$ operator $O_{ij}^{(\pm 2)}$
in terms of the polarization vectors as
\begin{eqnarray}
O_{ij}^{(\pm 2)}(\hat{\bm k})=-\sqrt{\frac{3}{8}}e^{(\pm )}_i(\hat{\bm k})e^{(\pm )}_j(\hat{\bm k})\,.
\end{eqnarray}
Since this operator obviously satisfies the transverse-traceless condition, the second-order tensor metric
perturbations can be expanded as
\begin{eqnarray}
h_{ij}(\eta ,{\bm x})=\int\frac{{\rm d}^3{\bm k}}{(2\pi )^3}\sum_{\sigma =\pm 2}h_\sigma (\eta ,{\bm k})O_{ij}^{(\sigma )}(\hat{\bm k})e^{-i{\bm k}\cdot{\bm x}}\,.
\end{eqnarray}
With these convention, we define the unequal-time power spectrum as
\begin{eqnarray}
\Braket{h_\sigma^* (\eta ,{\bm k})h_{\sigma'}(\eta' ,{\bm k}')}
=\left( 2\pi\right)^3\delta_{\sigma\sigma'}\delta^{3}_{\rm D}\left({\bm k}-{\bm k}'\right)\frac{1}{3} P_h(\eta ,\eta' ,k)\,.
\end{eqnarray}

Primordial gravitational waves are generated in the very early Universe and the representative sources for tensor mode.
Its effect on the CMB lensing and the shear measurement has been discussed in the literature \cite{Li:2006si,Dodelson:2003bv}.
For the evolution of primordial gravitational waves, we introduce the PGW transfer function $\mathcal{T}_h^{({\rm PGW})}(k\eta )$,
which basically describes its sub-horizon evolution. In terms of this, we can write the power spectrum as
\begin{eqnarray}
\frac{k^{3}}{2\pi^{2}}P_h^{({\rm PGW})}(\eta ,\eta' ,k)=r\Delta^2_{\mathcal{R}}(k_{0})\left(\frac{k}{k_0}\right)^{n_{\rm t}}
\mathcal{T}_h^{({\rm PGW})}(k\eta )\mathcal{T}_h^{({\rm PGW})}(k\eta' ) ~.
\end{eqnarray}
In our analysis we adopt $r=0.1$, $n_{\rm t}=0$ as the fiducial values, and use $\mathcal{T}^{({\rm PGW})}_h=3j_1(k\eta )/k\eta$ for simplicity.
The corrections due to the effects during the radiation-dominated era would be small and we neglect this small correction throughout this paper.

Similar to the case of the vector mode discussed in the previous section, the second-order source terms
induce the tensor metric perturbations, which are expected to be one of the possible sources of
the curl mode and B-mode shear signals~\cite{Sarkar:2008ii}.
The analytic model of the power spectrum for the second-order tensor mode induced
by the product of the first order scalar metric potentials has been discussed in \cite{Mollerach:2003nq,Baumann:2007zm}.
The approximate form of the power spectrum derived in \cite{Mollerach:2003nq} is given by
\begin{eqnarray}
\frac{k^{3}}{2\pi^{2}}P^{({\rm MHM})}_h(\eta ,\eta' ,k) 
=\frac{6}{25}C_{\rm T}\Delta^{4}_{\mathcal{R}}(k_{0})\left( \frac{k}{k_{*}}\right)^{-1}W_{\rm T}(k/k_{*})
\mathcal{T}^{({\rm MHM})}_h(k\eta)\mathcal{T}^{({\rm MHM})}_h(k\eta')~, \label{eq: 2nd order mollerach}
\end{eqnarray}
with $C_{\rm T} \approx 0.062$ and $W_{\rm T}(x) = \left( 1+7x+5x^{2}\right)^{-3}$.
The transfer function for the second-order tensor mode is 
\begin{eqnarray}
\mathcal{T}^{({\rm MHM})}_h(k\eta) &=& \left( 1-\frac{3j_{1}(k\eta)}{k\eta}\right) g^{2}_{\infty} ~.
\end{eqnarray}
The correction factor $g_{\infty}$ attributed to the effect of dark energy is defined as $g_{\infty}\equiv\lim_{z\to\infty}g(z) \approx 1.3136$.
We note that this formula is valid only after matter-radiation equality time.
While the correction during the radiation-dominated era for the second-order tensor mode has been considered in \cite{Baumann:2007zm},
for the purposes of comparing the second-order vector and tensor modes, 
we will neglect such correction since it must be small \cite{Sarkar:2008ii}.

%-----------------------------------------subsection
\subsection{Comparison of each model} \label{sec: preliminary}
\begin{figure}[t]
\begin{center}
\rotatebox{0}{\includegraphics[width=0.6\textwidth]{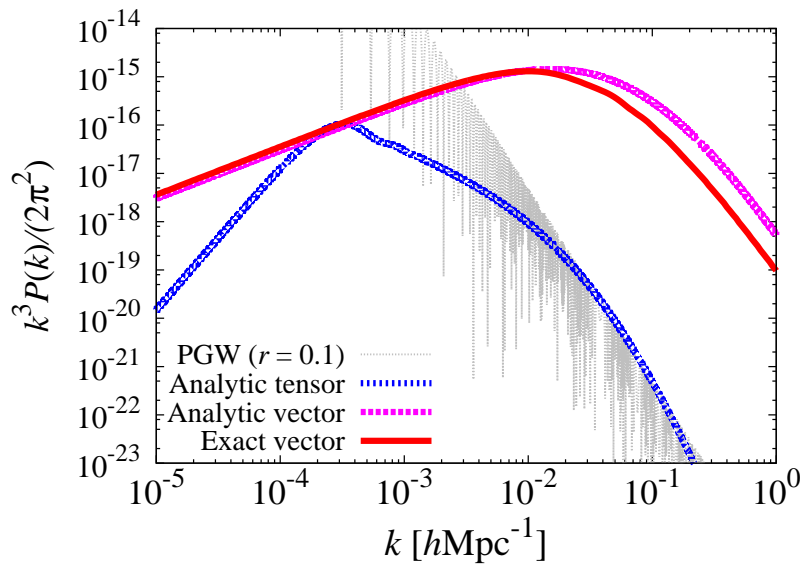}}
\end{center}
\caption{%
The power spectrum of the primordial gravitational waves with $r = 0.1$ (PGW), the second-order tensor mode (Analytic tensor), the analytical approximate solution 
of the second-order vector mode (Analytic vector), and the numerical solution of the second-order vector mode (Exact vector) at the present time ($1+z = 1$).
The second-order vector mode dominates on small scales rather than the second-order tensor mode.
The second-order vector mode derived by numerical calculation is slightly smaller than that derived by analytic approximation on smaller scales.
}
\label{fig: compare power}
\end{figure}

For comparison, the power spectra for these models at the present time are shown in Fig.~\ref{fig: compare power}.
%===
%===
The power spectrum for the second-order vector mode is larger than that for the second-order tensor mode.
Therefore, it is expected that the second-order vector mode induces a larger lensing signal than the second-order tensor mode on small scales.
Furthermore, on small scales, the second-order vector mode has a larger amplitude than the primordial gravitational waves with $r = 0.1$.
In other words, the second-order vector mode has the possibility of being detected by cosmological observations on small scales, unlike the primordial gravitational waves.
We explain the reason why the discrepancy between the exact vector model and the analytical vector model appears.
From Fig.~\ref{fig: compare power}, we can see that the amplitude of the exact model is smaller than that of the analytical vector model for $k \gtrsim k_{\rm eq}$, where $k_{\rm eq} \approx 10^{-2}~h{\rm Mpc}^{-1}$ is the horizon scale at the time of matter-radiation equality.
For $k \gtrsim k_{\rm eq}$, 
the analytical vector model does not consider the effect of the
small suppression around the matter-radiation equality time.  The small
suppression on small scales arises because the first order scalar potentials,
i.e., the source of the second order vector metric perturbations, decay so
rapidly in the radiation dominated era that the source on those scales can
sustain the vector perturbations and make them grow proportional to the scale
factor only after some time has passed since the
matter-radiation equality time (see the blue and magenta lines in Fig.~1).
This small suppression is not included in the analytical model.
The exact model is about ten times smaller than the analytical vector model for $k \gtrsim k_{\rm eq}$.

From the above discussions, we can understand when the suppression is determined.
The analytical vector model does not change its peak since this model is calculated in the flat $\Lambda$CDM model with matters and the cosmological constant.
Therefore, the difference between the analytical vector model and exact one is appeared until the matter-radiation equality.
In conclusion, the factor about $10$ suppression is determined at the matter-radiation equality time.

The tendency for the second-order vector mode is quite similar to that for the second-order tensor mode in Ref.~\cite{Sarkar:2008ii}.
In the next section, we will show the numerical results of the weak lensing induced by the second-order vector perturbation.
%-----------------------------------------------------------------------------------section
\section{weak lensing signals}
In this section, we present a short review of the full-sky formalism for the weak lensing induced by the vector and tensor modes following Refs.~\cite{2012JCAP...01..007N,2012JCAP...10..030Y,Yamauchi:2013fra}.
The weak lensing can roughly be classified into two observed objects.

First, the CMB photons emitted from the last scattering surface are deflected by the gravitational potentials related to the large-scale structure, which is called CMB lensing.
The CMB lensing is mainly caused by the scalar gravitational potential.
However, vector and tensor perturbations can also affect the deflection angle of photons, and the vector and tensor modes imprint characteristic deflection patterns on the CMB lensing.
The deflection angle of the CMB photons can be written as the gradient of the scalar potential (gradient-mode) and the rotation of the pseudo-scalar potential (curl-mode).
By using the nature of the parity these lensing potentials can be reconstructed independently,
even when the gradient mode dominates the CMB lensing signals~\cite{Namikawa:2011cs}.

Second, photons emitted from galaxies are lensed by the large-scale structure, causing the shapes of galaxies to be deformed.
This is known as cosmic shear.
By studying the deformation pattern statistically, we can distinguish traces of the scalar, vector, and tensor perturbations.
The deformation pattern of the shapes of galaxies can be decomposed into parity-even (E-mode) and parity-odd (B-mode) components.

In the following subsections, we present the full-sky formalism for the deflection angle and the deformation pattern, which are related to the geodesic equation and the Jacobi map, respectively.
Note that it is sufficient to work without the Hubble expansion since the geodesic equation is invariant under the conformal transformation.
%-----------------------------------------subsection
\subsection{Curl mode}
The projected deflection angle on the celestial sphere in the direction $\hat{\bm n}$, $\Delta_{a}(\hat{\bm n})$, 
is generally decomposed into the (parity-even) gradient and (parity-odd) curl modes, expressed as
\begin{equation}
\Delta_{a}(\hat{\bm n}) = \phi(\hat{\bm n})_{:a} + \varpi(\hat{\bm n})_{:b}\,\epsilon^{b}{}_{a} ~,
\end{equation}
where a colon denotes a covariant derivative on the unit sphere and 
$\epsilon^{b}{}_{a}$ is the covariant two-dimensional Levi-Civita tensor.
The quantities $\phi$ and $\varpi$ denote the gradient- and
curl-modes of the deflection angle, respectively.
Raising or lowering indices of the two-sphere vector are done by $\omega_{ab}$ defined in Eq.~(\ref{eq: metric spherical}).
In our analysis, we focus only on the curl-mode as the signal of the CMB lensing.
We will expand the curl-mode by the spherical harmonics since we observe the scalar and pseudo-scalar lensing potentials on the celestial sphere:
\begin{equation}
\varpi(\hat{\bm n}) = \sum_{\ell, m} \varpi_{\ell ,m}Y_{\ell ,m}(\hat{\bm n}).
\end{equation}
With the coefficients of the harmonics, the angular power spectrum for the curl-mode is defined by
\begin{equation}
C^{\varpi\varpi}_{\ell} = \frac{1}{2\ell + 1}\sum_{m = -\ell}^{\ell} \Braket{\varpi^{*}_{\ell ,m}\varpi_{\ell ,m}} ~. \label{eq: Cl real space}
\end{equation}

The deflection angle of the light path obeys the null geodesic equation in the perturbed Universe.
Solving the spatial parts of geodesic equation for the photon rays at the first-order, we obtain the explicit expression of 
the curl-mode in terms of the metric perturbations as \cite{Yamauchi:2013fra}
\begin{eqnarray}
\varpi (\hat{\bm n})^{:a}{}_{:a} &=& -\int^{\chi_{\rm S}}_{0}{\rm d}\chi \frac{\chi_{\rm S}-\chi}{\chi_{\rm S}\chi}
\left\{ \frac{\rm d}{{\rm d}\chi}\Bigl( \chi \Omega^{a}(\eta_0 -\chi ,\chi\hat{\bm n}){}_{:b}\,\epsilon^{b}{}_{a}\Bigr)\right\} ~, \label{eq: der varpi}
\end{eqnarray}
where $\chi_{\rm S}$ is the comoving distance of the last scattering surface from the observer and
$\Omega_a$ in the Poisson gauge is written only in terms of the projected vector and tensor modes:
\begin{eqnarray}
\Omega_{a}(\eta_0 -\chi ,\chi\hat{\bm n}) &\equiv& \Bigl\{ -\sigma_{i}(\eta_0 -\chi ,\chi\hat{\bm n})+h_{ij}(\eta_0 -\chi , \chi\hat{\bm n})\hat n^j\Bigr\}\, e^{i}_{a}(\hat{\bm n})~.
\end{eqnarray}
Here we used the Born approximation in Eq.~\eqref{eq: der varpi} and the basis vectors in real space, $\hat{\bm n}$ and ${\bm e}_a(\hat{\bm n})$, 
have been defined in Eqs.~\eqref{eq: basis vectors1}-\eqref{eq: basis vectors3}.
We can obviously see that the curl-mode is generated by the vector and/or tensor metric perturbations, as already mentioned in the Introduction. 

The angular power spectra induced by the vector and tensor modes, respectively, are expressed in terms of the unequal-time power spectra as~\cite{Yamauchi:2013fra}
\begin{widetext}
\begin{eqnarray}
&&C^{\varpi\varpi ,({\rm v})}_\ell =\frac{\pi}{2}\int^\infty_0 k^2{\rm d}k
\int^{\chi_{\rm S}}_0{\rm d}\chi\int^{\chi_{\rm S}}_0{\rm d}\chi'\,
\mathcal{S}^{({\rm v})}_{\varpi ,\ell}(k,\chi )\mathcal{S}^{({\rm v})}_{\varpi ,\ell}(k,\chi' )
P_\sigma (\eta_0 -\chi ,\eta_0 -\chi' ,k)
\,, \label{eq: curl from vector}\\
&&C^{\varpi\varpi ,({\rm t})}_\ell =\frac{\pi}{2}\int^\infty_0 k^2{\rm d}k
\int^{\chi_{\rm S}}_0{\rm d}\chi\int^{\chi_{\rm S}}_0{\rm d}\chi'\,
\mathcal{S}^{({\rm t})}_{\varpi ,\ell}(k,\chi )\mathcal{S}^{({\rm t})}_{\varpi ,\ell}(k,\chi' )
P_h (\eta_0 -\chi ,\eta_0 -\chi' ,k)
\,, \label{eq: curl from tensor}
\end{eqnarray}
\end{widetext}
where $S_{\varpi ,\ell}^{({\rm v,t})}$ are the weight function for the vector and tensor modes,
\begin{eqnarray}
\mathcal{S}^{({\rm v})}_{\varpi, \ell}(k,\chi) &=& \sqrt{\frac{(\ell -1)!}{(\ell +1)!}}\frac{j_{\ell}(k\chi)}{\chi}~, \\
\mathcal{S}^{({\rm t})}_{\varpi, \ell}(k,\chi) &=& \frac{1}{2}\frac{(\ell -1)!}{(\ell +1)!}\sqrt{\frac{(\ell +2)!}{(\ell -2)!}}\frac{j_{\ell}(k\chi)}{k\chi^2} ~, \label{eq: curl weight from tensor}
\end{eqnarray}
and $P_{\sigma}$ and $P_{h}$ in Eqs.~(\ref{eq: curl from vector}) and (\ref{eq: curl from tensor}) are the power spectra of the vector and tensor metric perturbations presented in Sec.~{\ref{sec: second-order}}, respectively.

%-----------------------------------------subsection
\subsection{B-mode shear}
In this subsection, we summarize the formalism for the cosmic shear described by the Jacobi map, which maps the intrinsic light bundle to the observed light bundle.
The geodesic deviation equation is required to handle the deformation of the light bundle \cite{Seitz:1994xf}.
The deviation vector projected on the celestial sphere $\xi^{a}$ obeys \cite{Seitz:1994xf}
\begin{equation}
\frac{{\rm d}^{2}\xi^{a}}{{\rm d}\chi^{2}} = \mathcal{T}^{a}{}_{b}\,\xi^{a} ~, \label{eq: deviation eq}
\end{equation}
where we define the perturbed symmetric optical tidal matrix $\mathcal{T}^{a}{}_{b}$,
which is related to the deformation of the light bundle.
Its explicit form is given by
\begin{equation}
\mathcal{T}^{a}{}_{b} = -R_{\mu\alpha\nu\beta}\frac{{\rm d}x^{\mu}}{{\rm d}\chi}\frac{{\rm d}x^{\nu}}{{\rm d}\chi}e^{\alpha a}e^{\beta}_{b} ~.
\end{equation}
We set the initial conditions in the observer's frame of reference, $\left. \xi^{a}\right|_{\chi =0} = 0$ and $\left. {\rm d}\xi^{a}/{\rm d}\chi\right|_{\chi =0} = \delta\theta^{a}_{0}$, 
and introduce the Jacobi map as
\begin{equation}
\xi^{a} = \mathcal{D}^{a}{}_{b}\, \delta\theta^{b}_{0} ~.
\end{equation}
This matrix characterizes the deformation of light bundle.
Substituting this relation into Eq.~(\ref{eq: deviation eq}),
the Jacobi map satisfies the following equation:
\begin{equation}
\frac{{\rm d}^{2}\mathcal{D}^{a}{}_{b}}{{\rm d}\chi^{2}} = \mathcal{T}^{a}{}_{c}\,\mathcal{D}^{c}{}_{b} ~, \label{eq: jacobi 1}
\end{equation}
where the initial conditions for the Jacobi map are $\left. \mathcal{D}^{a}{}_{b} \right|_{\chi =0} = 0$ and $\left. {\rm d}\mathcal{D}^{a}{}_{b}/{\rm d}\chi\right|_{\chi =0} = \delta^{a}{}_{b}$.
From here, we decompose the Jacobi map into the spin-0 and spin-2 variables as
\begin{equation}
{}_{0}\mathcal{D} = \mathcal{D}_{ab}e^{a}_{(+)}e^{b}_{(-)} ~,~~~ {}_{\pm 2}\mathcal{D} = \mathcal{D}_{ab}e^{a}_{(\pm )}e^{b}_{(\pm )} ~,
\end{equation}
where we have introduced the polarization basis with respect to a two-dimensional vector on the sky as
$e^{a}_{(\pm )} (\hat{\bm n})= e^{i}_{(\pm )}(\hat{\bm n}) e^{a}_{i}(\hat{\bm n})$.
Note that the above polarization basis is in terms of spin-1 variables, namely, $e^{a}_{(\pm)} \to \exp\left(\pm i\alpha\right)e^{a}_{(\pm)}$ under the rotation around $\hat{\bm n}$ by an angle $\alpha$.
Furthermore, we define the reduced shear by using the Jacobi map as
\begin{equation}
g = -\frac{{}_{+2}\mathcal{D}}{{}_{0}\mathcal{D}} ~,~~~ g^{*} = -\frac{{}_{-2}\mathcal{D}}{{}_{0}\mathcal{D}} ~. \label{eq: reduced shear Jacobi}
\end{equation}
With these we introduce the E- and B-modes for the reduced shear fields and present the full-sky formalism
for weak lensing measurements.
We first expand the reduced shear, which is a spin-2 variable, by the spin-2 spherical harmonics as
\begin{eqnarray}
g(\hat{\bm n}) &=& \sum_{\ell, m} \left( E_{\ell m}+ iB_{\ell m}\right) {}_{+2}Y_{\ell m}(\hat{\bm n}) ~, \label{eq: def E and B 1}\\
g^{*}(\hat{\bm n}) &=& \sum_{\ell, m} \left( E_{\ell m}- iB_{\ell m}\right) {}_{-2}Y_{\ell m}(\hat{\bm n}) ~, \label{eq: def E and B 2}
\end{eqnarray}
where $E_{\ell m}$ and $B_{\ell m}$ have electric and magnetic parities, i.e., $(-1)^{\ell}$ and $(-1)^{\ell + 1}$, and 
are called the E- and B-modes, respectively.
The angular power spectrum of these modes is defined as
\begin{equation}
C^{\rm XX'}_{\ell} = \frac{1}{2\ell + 1} \sum^{\ell}_{m = -\ell}\Braket{X^{*}_{\ell m}X'_{\ell m}} ~,
\end{equation}
where $X$ and $X'$ take $E$ or $B$.
Note that the scalar, vector, and tensor modes can induce the E-mode.
On the other hand, only the vector and tensor modes can induce the B-mode.

To obtain the expression relevant for the weak lensing measurements, let us expand Eq.~\eqref{eq: jacobi 1}
and solve it order by order. 
Since the tidal matrix vanishes in unperturbed spacetime, the zeroth-order solution of Jacobi map
trivially reduces to ${\cal D}^{(0)}{}^a{}_b=\chi_{\rm S}\delta^a{}_b$. 
Substituting this into the first-order geodesic deviation equation, we obtain 
the symmetric trace-free part of the Jacobi map up to the next-leading order~\cite{Yamauchi:2013fra}:
\begin{eqnarray}
\frac{1}{\chi_{\rm S}}{\cal D}_{\Braket{ab}} =
\int^{\chi_{\rm S}}_{0}{{\rm d}\chi}\frac{\chi_{\rm S}-\chi}{\chi_{\rm S}\chi}
\left\{ \Upsilon (\eta_0 -\chi ,\chi\hat{\bm n})_{:\Braket{ab}} -\frac{\rm d}{{\rm d}\chi}\left( \chi\Omega_{\langle a}(\eta_0 -\chi ,\chi\hat{\bm n})_{:b\rangle}\right)\right\}
+\frac{1}{2}\bigl[ h_{\langle ab\rangle}\bigr]^{\chi_{\rm S}}_{0}~,
\label{eq:Jacobi map solution}
\end{eqnarray}
where $h_{ab} \equiv h_{ij}e^{i}_{a}e^{j}_{b}$, $[f]^{\chi_{\rm S}}_0 = f(\eta_0 -\chi_{\rm S},\chi_{\rm S}\hat{\bm n}) - f(\eta_0 ,{\bm 0})$,
and $\Upsilon$ is written in terms of the metric perturbations as
\begin{eqnarray}
\Upsilon =-\left(\Psi +\Phi\right) -\sigma_i\hat n^i+\frac{1}{2}h_{ij}\hat n^i\hat n^j\,.
\end{eqnarray}
Once multiplying the resultant Jacobi map \eqref{eq:Jacobi map solution} by $e^a_{(+)}e^b_{(+)}$ and $e^a_{(-)}e^b_{(-)}$, 
we can obtain the explicit expression for the spin-$+2$ and $-2$ reduced shear fields in terms of the metric perturbations.

In this paper, we focus on the B-mode shear induced by the vector and tensor modes.
Note that we consider galaxy observations such as imaging surveys, which does not accurately divide the redshift of each galaxy.
To discuss the weak lensing measurements from imaging surveys, the redshift distribution of background galaxies $N(\chi )$ 
should be taken into account.

The angular power spectrum for the B-mode shear is written in terms of the weight function
and the unequal-time power spectrum for the vector and tensor modes,
\begin{eqnarray}
&&C^{{\rm BB} ,({\rm v})}_\ell =\frac{\pi}{2}\int^\infty_0 k^2{\rm d}k
\int^{\chi_{\rm S}}_0{\rm d}\chi\int^{\chi_{\rm S}}_0{\rm d}\chi'\,
\mathcal{S}^{({\rm v})}_{{\rm B} ,\ell}(k,\chi )\mathcal{S}^{({\rm v})}_{{\rm B} ,\ell}(k,\chi' )
P_\sigma (\eta_0 -\chi ,\eta_0 -\chi' ,k)
\,, \label{eq: B from vector}\\
&&C^{{\rm BB} ,({\rm t})}_\ell =\frac{\pi}{2}\int^\infty_0 k^2{\rm d}k
\int^{\chi_{\rm S}}_0{\rm d}\chi\int^{\chi_{\rm S}}_0{\rm d}\chi'\,
\mathcal{S}^{({\rm t})}_{{\rm B} ,\ell}(k,\chi )\mathcal{S}^{({\rm t})}_{{\rm B} ,\ell}(k,\chi' )
P_h (\eta_0 -\chi ,\eta_0 -\chi' ,k)
\,, \label{eq: B from tensor}
\end{eqnarray}
where the weight functions for the B-mode shear, $S_{{\rm B} ,\ell}^{({\rm v,t})}$, are given by
\begin{widetext}
\begin{eqnarray}
\mathcal{S}^{({\rm v})}_{{\rm B},\ell} &=& \frac{1}{2}\sqrt{\frac{(\ell + 2)!(\ell -1)!}{(\ell - 2)!(\ell +1)!}}
\int^{\infty}_{\chi}{\rm d}\chi_{\rm S}\frac{N(\chi_{\rm S})}{N_{\rm g}}\frac{j_{\ell}(k\chi)}{\chi} ~, \\
\mathcal{S}^{({\rm t})}_{{\rm B},\ell} &=&\frac{1}{4}\sqrt{\frac{(\ell + 2)!(\ell -1)!}{(\ell - 2)!(\ell +1)!}}
\int^{\infty}_{\chi}{\rm d}\chi_{\rm S}\frac{N(\chi_{\rm S})}{N_{\rm g}}
\frac{j_{\ell}(k\chi)}{k\chi^2} 
- \frac{1}{4}\frac{N(\chi )}{N_{\rm g}}
\left( j'_{\ell}(k\chi) + 2\frac{j_{\ell}(k\chi)}{k\chi}\right) ~, \label{eq: B weight from tensor}
\end{eqnarray}
\end{widetext}
and $P_{\sigma}$ and $P_{h}$ in Eqs.~(\ref{eq: B from vector}) and (\ref{eq: B from tensor}) are the power spectra of the vector and tensor metric perturbations presented in Sec.~{\ref{sec: second-order}}, respectively.
In our calculation, we assume a distribution of galaxies $N(\chi)$, which can usually be taken to be (see, e.g., Ref.~\cite{Yamamoto:2007gd})
\begin{equation}
N(\chi_{\rm S}){\rm d}\chi_{\rm S} = N_{\rm g}\frac{3}{2}\frac{z^{2}_{\rm S}}{(0.64 z_{\rm m})^{3}}\exp\left[ -\left( \frac{z_{\rm S}}{0.64z_{\rm m}}\right)^{3/2}\right]{\rm d}z_{\rm S} ~,
\end{equation}
where $z_{\rm m}$ is the mean redshift,
and the number of galaxies per square arc-minute $N_{\rm g}$ is defined as
\begin{equation}
N_{\rm g} \equiv \int^{\infty}_{0}{\rm d}\chi N(\chi) ~.
\end{equation}
In this paper, we focus on four survey designs: DES~\cite{Abbott:2005bi}, HSC~\cite{HSC:coll}, SKA~\cite{Brown:2015ucq}, and LSST~\cite{2009arXiv0912.0201L}.
The experimental specifications of each survey design are summarized in Table.~\ref{tab: designs}.
\begin{table}[htb]
{\tabcolsep = 3mm
\begin{tabular}{| c || c | c | c |} \hline
 & $f_{\rm sky}$ & $z_{\rm m}$ & $N_{\rm g} {\rm [arcmin^{-2}]}$ \\ \hline
DES & $0.125$ & $0.5$ & $12$ \\ \hline
HSC & $0.05$ & $1.0$ & $35$ \\ \hline
SKA & $0.75$ & $1.6$ & $10$ \\ \hline
LSST & $0.5$ & $1.5$ & $100$ \\ \hline
\end{tabular}}
\caption{The experimental specifications of DES, HSC, SKA, and LSST.
It is shown that the sky coverage $f_{\rm sky}$, the mean redshift $z_{\rm m}$, and the number of the galaxies per square arc minute $N_{\rm g}$.}
\label{tab: designs}
\end{table}

%-----------------------------------------------------------------------------------section
\section{Weak lensing induced by second-order vector mode}
In this section, we show our main results and discuss the size of the effect of second-order vector modes.
We now calculate the weak lensing signals from the second-order vector mode by performing the numerical calculation
(hereafter referred to as the exact vector).
For comparison, the results for the signals from the primordial gravitational waves with $r = 0.1$, 
the second-order tensor mode (analytic tensor), 
the second-order vector mode (analytic vector) are also shown.

\begin{figure}[t]
\begin{center}
\rotatebox{0}{\includegraphics[width=0.7\textwidth]{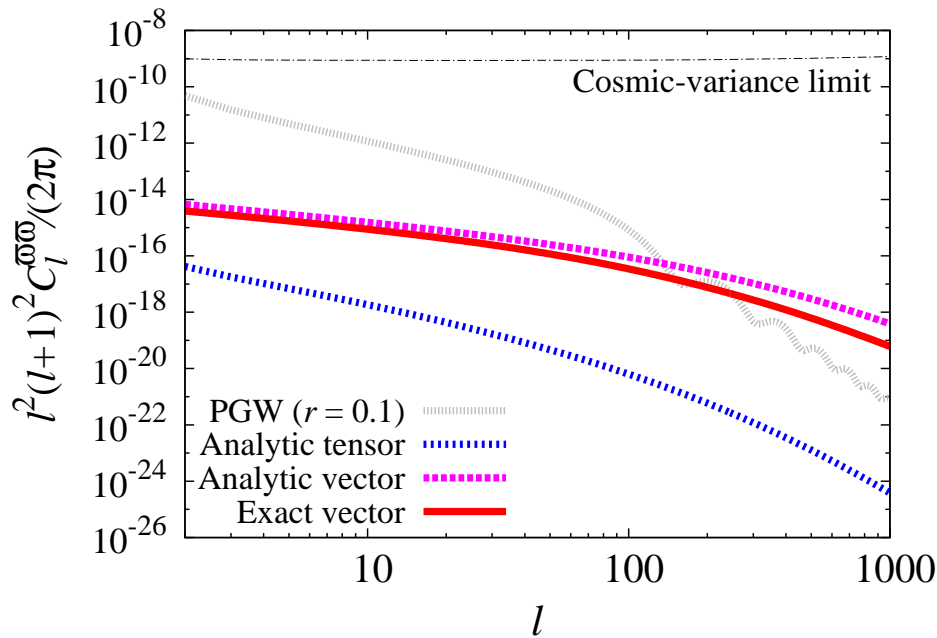}}
\end{center}
\caption{%
The angular power spectrum of the weak lensing curl-mode.
As we expected in section \ref{sec: preliminary}, the second-order vector mode dominates on small scales.
Furthermore, the second-order tensor mode becomes a sub-dominant contribution to the weak lensing curl-mode.
The expected noise from the cosmic variance limit is also shown.
}
\label{fig: result curl mode}
\end{figure}

%curl-mode
First, we show the angular power spectrum of the curl-mode in Fig.~\ref{fig: result curl mode} for the CMB lensing measurement.
The CMB lensing reconstruction technique can decompose the lensing potential into the gradient and curl modes.
Although the gradient mode dominates the lensing signals, owing to this technique, the information about the gradient and curl modes can be extracted independently.
Even when we neglect the instrumental noise, we need to take into account for the reconstruction noise only.
The noise estimated by the ideal CMB weak lensing measurement is determined by a cosmic-variance limited reconstruction of the curl-mode \cite{Cooray:2005hm,2012JCAP...01..007N}.
%===
%===
We found that the curl-mode induced by the primordial gravitational waves dominates on large scales, $\ell \lesssim 200$, while that by the second-order vector mode dominates on small scales, $\ell \gtrsim 200$.
As seen in Fig.~\ref{fig: compare power}, the power spectrum for the vector mode has a peak at the scale corresponding to matter-radiation equality.
On the other hand, those for the primordial and second-order tensor modes have their peaks at the horizon scales.
Therefore, the second-order vector mode can affect smaller scales than the primordial or the second-order tensor mode does.
As expected, the second-order tensor mode gives a subdominant contribution to the weak lensing curl-mode.
This feature is similar to the CMB polarization anisotropy \cite{Mollerach:2003nq} and the weak lensing gradient-mode \cite{Andrianomena:2014sya}.

However, unfortunately, even if we consider ideal experiments, i.e., only the cosmic-variance limited error, the weak lensing curl-mode signals do not exceed the expected noise.
Although the curl-mode induced by the second-order vector mode dominates the signal of the curl-mode on small scales, it will be difficult to detect the second-order vector and tensor weak lensing signals in future experiments.
We conclude that the curl-mode induced by the second-order modes cannot be detected by any CMB observations in the future because of the cosmic variance limit.
On the other hand, recently, a new possibility has emerged of detecting the weak lensing signals in 21cm observations \cite{2010PhRvL.105p1302M,2012PhRvL.108u1301B}.
The angular power spectrum of 21cm fluctuations can be expanded up to $\ell \sim 10^{7}$ since they do not have diffusion scales unlike CMB fluctuations.
Furthermore, the 21cm fluctuations enable us to observe the fluctuation at different frequencies which corresponding to the different distances.
Therefore, the signal-to-noise ratio can be substantially improved.
For example, in Ref.~\cite{2012PhRvL.108u1301B}, the observable scalar-to-tensor ratio reaches $r \approx 10^{-9}$.
If this sensitivity is reached in the future observations, the 21cm curl-mode induced by the second-order vector mode should be detected.
The 21cm fluctuations would be a good probe of the weak lensing curl-mode.

\begin{figure}[t]
\begin{center}
\rotatebox{0}{\includegraphics[width=0.45\textwidth]{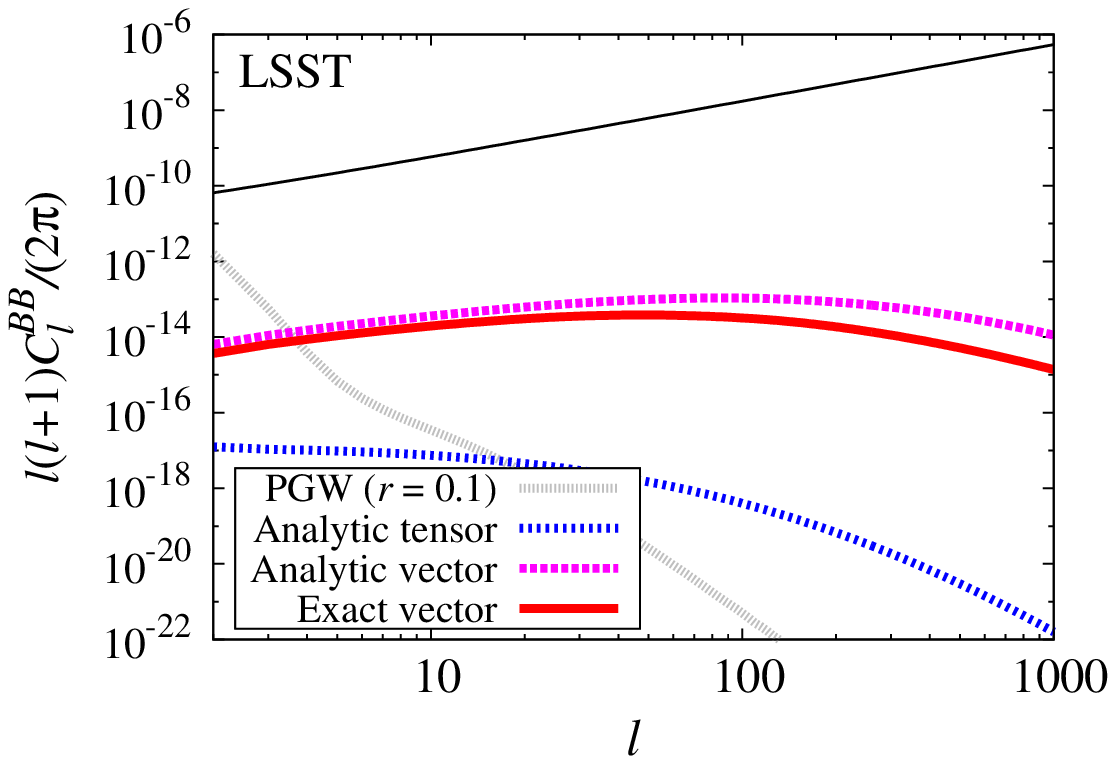}}
\rotatebox{0}{\includegraphics[width=0.45\textwidth]{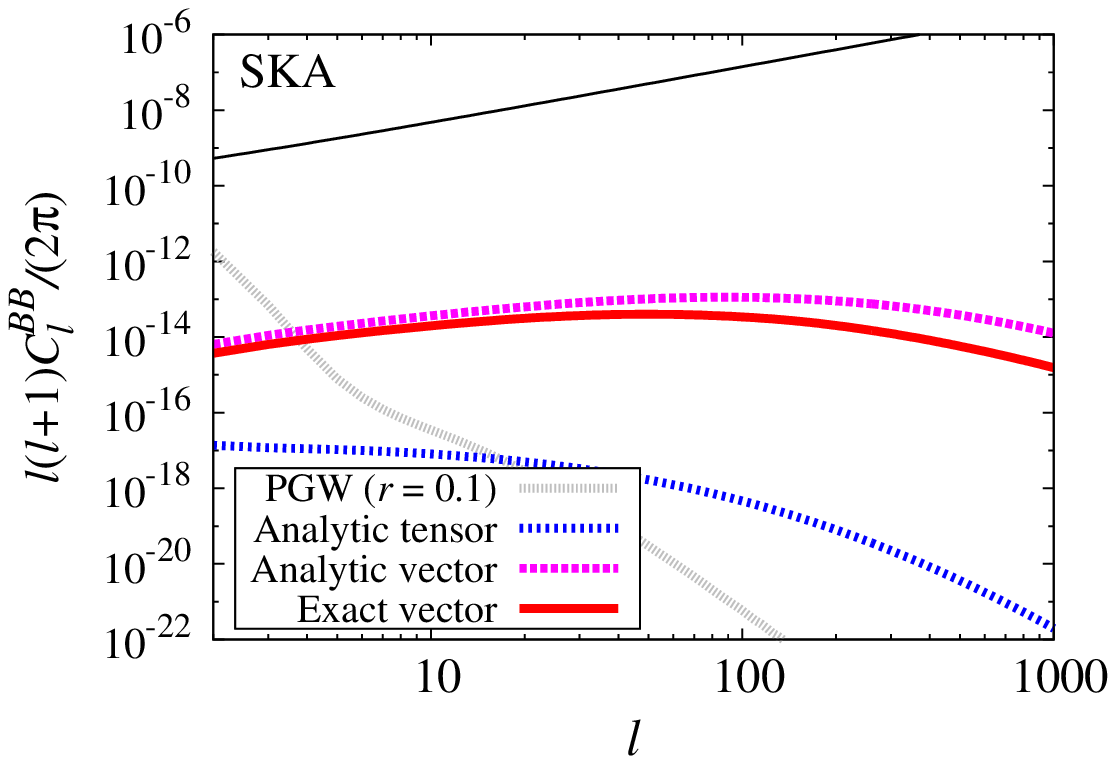}}
\rotatebox{0}{\includegraphics[width=0.45\textwidth]{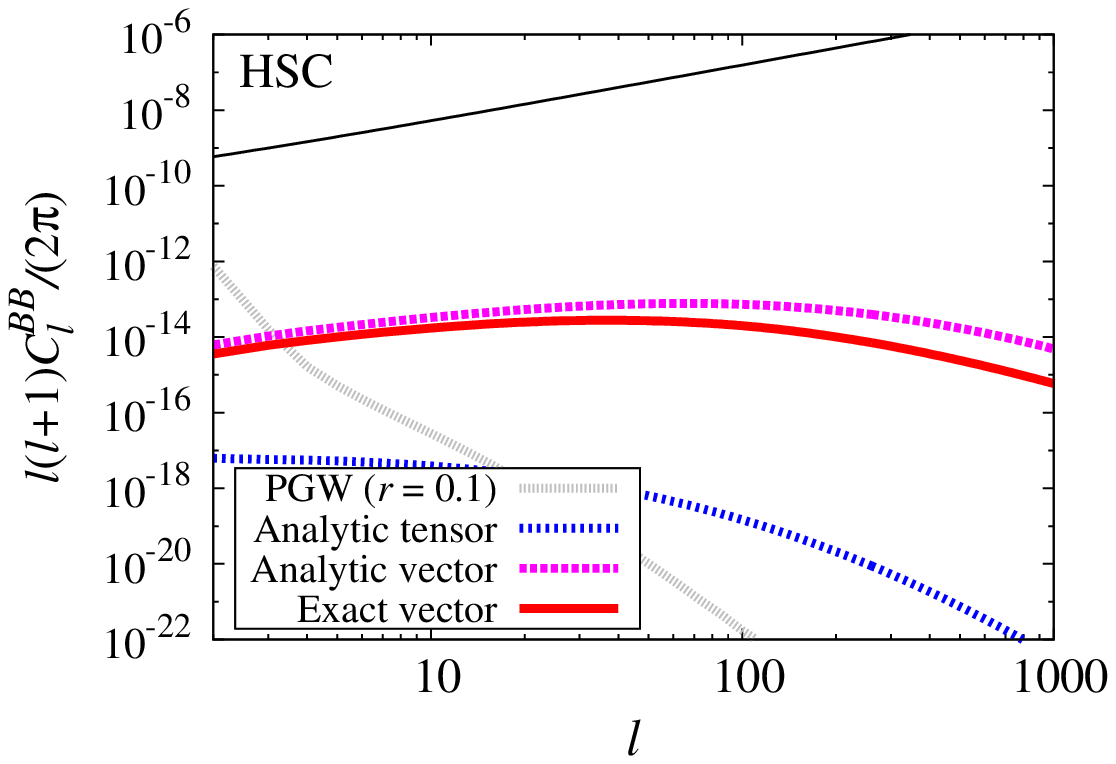}}
\rotatebox{0}{\includegraphics[width=0.45\textwidth]{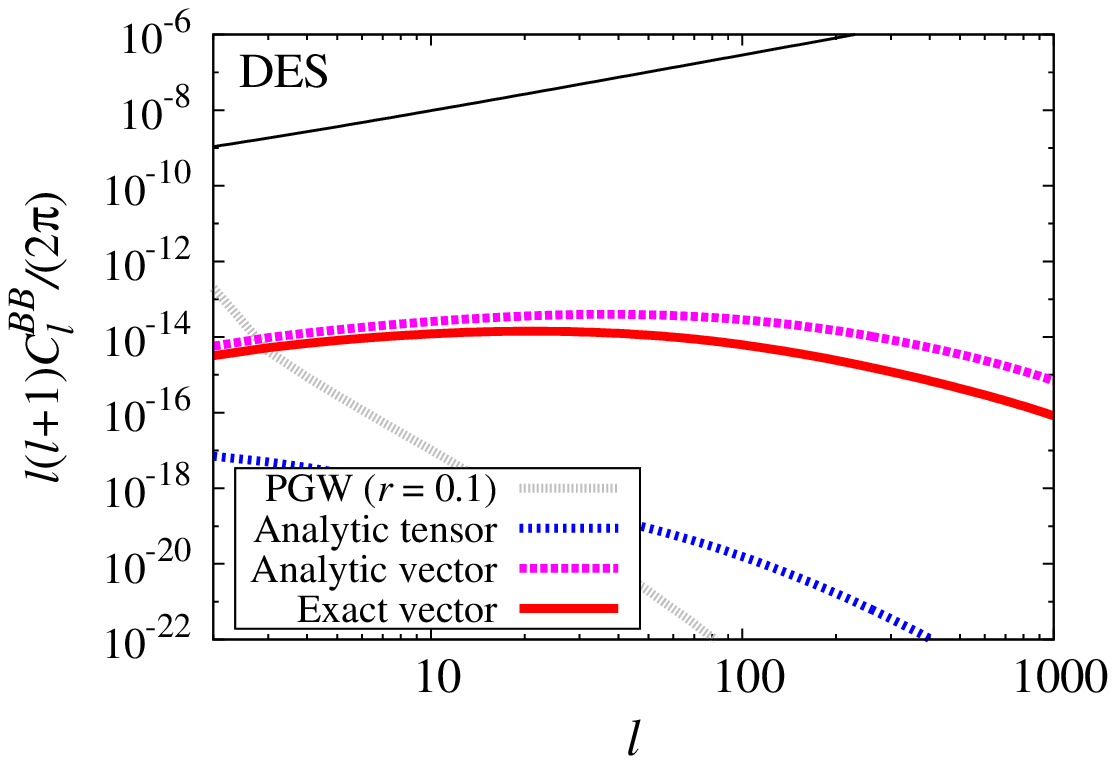}}
\end{center}
\caption{%
A angular power spectra of the weak lensing B-mode assumed four survey designs: LSST (top left), SKA (top right), HSC (bottom left), and DES (bottom right).
The second-order vector mode dominates the expected signals on small scales.
The black solid line shows the expected statistical error for each survey.
}
\label{fig: result BB}
\end{figure}

%B-mode
Second, we show the angular power spectrum of the B-mode shear with the four representative imaging surveys, DES, HSC, SKA, and LSST, in Fig.~\ref{fig: result BB}.
Unlike the CMB lensing, the statistical error in the cosmic shear measurements is determined by the intrinsic ellipticity of each galaxy.
In this paper, we assume that the error mainly originates from the intrinsic ellipticity of each galaxy as
\begin{equation}
N^{\rm BB}_{\ell} = \sqrt{\frac{2}{(2\ell + 1)f_{\rm sky}}}\frac{\Braket{\gamma^{2}_{\rm int}}}{3600N_{\rm g}(180/\pi)^{2}} ~, \label{eq: noise BB}
\end{equation}
where $\Braket{\gamma^{2}_{\rm int}}^{1/2}$ is the root-mean-square ellipticity of galaxies.
In this paper, we set $\Braket{\gamma^{2}_{\rm int}}^{1/2} = 0.3$ derived in Ref.~\cite{Bernstein:2001nz}.
The error in the cosmic shear measurements is mainly controlled by the sky coverage $f_{\rm sky}$ and the number of the galaxies per square arc minute $N_{\rm g}$ and we show the error expected by four survey designs in Table.~\ref{tab: designs}.
%===
%===
From Fig.~\ref{fig: result BB}, we can see that the B-mode induced by the second-order vector mode dominates on all scales except for the largest scale.
However, as is the case with the CMB lensing curl-mode, the cosmic shear B-mode induced by the second-order vector mode does not exceed the expected noise for each survey design.
From Eq.~(\ref{eq: noise BB}) and Fig.~\ref{fig: result BB}, the combined survey design parameter appeared in Eq.~(\ref{eq: noise BB}), i.e., $\sqrt{f_{\rm sky}}\times N_{\rm g}$, should be improved about $10^{4}$ compared with LSST to detect the B-mode signal.
Such an ultimate survey is quite unrealistic even in the distant future in contrast with the 21cm lensing observations.

We note that in this paper, we focus on the standard cosmological model which can characterize the primordial power spectrum by the primordial amplitude $\Delta^{2}_{\mathcal{R}}$ and the spectral index $n_{s}$ in Eq.~(\ref{eq: primordial power}).
However, the non-standard model may enhance the primordial power spectrum on smaller scales (e.g., \cite{Kawasaki:2006zv,Chluba:2012we}).
The second-order signals are sensitive to the enhancement on smaller scales since the mode mixing is introduced by the convolution of the small- and large-scale fluctuations.
The second-order signals would be useful to probe the small-scale physics related to the inflation model.

Let us consider the difference between the weak lensing induced by the second-order vector and tensor modes.
The equation of motion for the tensor metric perturbation has the form of a wave equation.
Therefore, the second-order tensor mode induced by the products of the first-order scalar modes cannot be amplified when the source remains constant 
in the matter-dominated era on sub-horizon scales~\cite{Saga:2014jca}.
On the other hand, the evolution of the vector metric perturbation is equivalent to that of the vorticity.
We can see that the vorticity with the source is well amplified in Eq.~(\ref{eq: sigma integrated}) even in such an era.
Therefore, the amplitude of the second-order vector mode is larger than that of the second-order tensor mode.

To conclude this section, we remark on other second-order contributions to the weak lensing curl- and B- modes.
During photon propagation, there are some corrections to the weak lensing formula induced by the geodesic effect \cite{2010PhRvD..81h3002B,2012PhRvD..86b3001B}.
The geodesic effect would have the possibility to enhance the curl- and B-mode signals.
However, this geodesic effect is induced not by the vector and tensor modes but by the product of the first-order scalar perturbations such as the Weyl potential, which we leave for future work.
%-----------------------------------------------------------------------------------section
\section{Summary}
In this paper, we explored the weak lensing signals induced by the second-order vector perturbation.
The weak lensing effects are classified into two observables: CMB lensing and cosmic shear.
Both the signals of the CMB lensing and cosmic shear can be decomposed into two modes by using parity, namely, the gradient- and curl-modes for the CMB lensing and the E- and B-modes for the cosmic shear.
The curl- and B-modes are only induced by the vector and tensor modes.
In the standard cosmology, the vector mode is neglected and the source of the curl- and B-modes is limited to the case of primordial gravitational waves, which have not been observed yet.
However, when we expand the cosmological perturbation theory up to the second order, the vector and tensor modes are naturally induced by the product of the first-order scalar perturbations.
As the first-order scalar perturbation theory is well established by a number of observations, the second-order vector and tensor modes do not include free parameters and are well determined.

We presented the effect of the second-order vector mode on the weak lensing for the first time.
The weak lensing induced by the second-order vector mode dominates on smaller scales rather than the primordial gravitational waves with $r = 0.1$ and the second-order tensor mode.
In particular, the cosmic shear induced by the second-order vector mode dominates on almost all scales.
This is because the second-order vector mode can be enhanced when the source exists in the matter-dominated epoch while the second-order tensor mode remains constant even if the source exists.
This difference also affects cosmological signatures such as the CMB polarization anisotropy.
However, the weak lensing signals induced by the second-order vector mode cannot exceed the expected noise estimated by the cosmic-variance limit and the shot-noise for the CMB lensing and cosmic shear, respectively.
Therefore, unfortunately, it seems difficult to detect the CMB curl- and B-modes induced by not only the second-order tensor mode but also the vector mode in the ongoing and forthcoming weak lensing observations.
However, the 21cm observations can decrease the expected noise and it may be possible that the 21cm lensing observations can be detect the 21cm lensing curl-mode.

Throughout this paper, we assume the standard cosmological model.
In other words, the primordial power spectrum is characterized by the amplitude and the spectral index.
However, non-standard cosmological models can enhance the primordial power on much smaller scales.
The weak lensing curl- and B-modes would become the good probe to search for the small-scale power spectrum and we leave this to future work.

%======================= Acknowledgements =======================%
\begin{acknowledgments}
One of the authors (S.S.) thanks Masato I.N. Kobayashi for useful discussions on the basics of weak lensing.
This work was supported in part by a Grant-in-Aid for JSPS Research under Grants No.~26-63 (S.S.) and No.~25-9800 (D.Y.)
and a JSPS Grant-in-Aid for Scientific Research under Grant No. 24340048 (K.I.).
We also acknowledge the Kobayashi-Maskawa Institute for the Origin of Particles and the Universe, Nagoya University, for providing useful computing resources for conducting this research.
\end{acknowledgments}
%======================= Bibtex =======================%
\bibliography{ref}
\end{document}